# Development of Photogrammetric Methods of Stress Analysis and Quality Control

*Donna L. Kubik, Northern Illinois University, DeKalb, IL, USA*
*John A. Greenwood, Fermi National Accelerator Laboratory, Batavia, IL, USA*
(in association with the MuCool Absorber Collaboration)

## 1. ABSTRACT

A photogrammetric method of stress analysis has been developed to test thin, nonstandard windows designed for hydrogen absorbers, major components of a muon cooling channel. The purpose of the absorber window tests is to demonstrate an understanding of the window behavior and strength as a function of applied pressure. This is done by comparing the deformation of the window, measured via photogrammetry, to the deformation predicted by finite element analysis (FEA). FEA analyses indicate a strong sensitivity of strain to the window thickness. Photogrammetric methods were chosen to measure the thickness of the window, thus providing data that are more accurate to the FEA. This, plus improvements made in hardware and testing procedures, resulted in a precision of 5 microns in all dimensions and substantial agreement with FEA predictions.

## 2. INTRODUCTION

Muon accelerators are being considered to extend the energy reach of high energy physics. Present approaches using hadron and electron colliders are reaching critical size and performance constraints. Hadron colliders are performance-constrained in that complicated many-particle collisions with a rapidly diminishing fraction (in numbers and energy) of the interactions in point-like new particle-state-production is produced. Lepton colliders produce simple interactions, and this magnifies the effective energy of collisions by more than an order of magnitude over hadron colliders. Extension of e+e- colliders to multi-Tev energies is constrained by 'beamstrahlung' (Appendix A) and synchrotron radiation effects, which increase as $(E_e / m_e)^4$. Muons, however, have negligible radiation and 'beamstrahlung' [1].

The liability of muons is that they decay with a lifetime of $2.2\text{x}10^{-6} E_\mu/m_\mu \text{s}$ and that they are created through decay into a diffuse phase space. The phase space can be reduced by ionization cooling, and the lifetime is sufficient for storage ring collisions. In ionization cooling, muons lose energy through ionization interactions while passing through material (referred to as absorber), losing both longitudinal and transverse momentum, and are reaccelerated, regaining only longitudinal momentum. The loss of transverse momentum reduces particle emittances, cooling the beam. However, the random process of multiple scattering in the material increases the RMS of the beam divergence, adding a heating term, which must be controlled in the cooling design [1].

The choice of absorber material is driven by a tradeoff between energy loss and multiple scattering in the material. Energy loss depends on Z/A, while scattering is a function of $Z^2$, indicating the best choice of material is hydrogen. Similarly, the container for the hydrogen should be made from a low-Z material. Beryllium would be optimal, but due to the toxic effects of beryllium, aluminum has been chosen. The container is a cylindrical vessel axially symmetric with the beam. The end-caps (referred to as windows) are designed to minimize thickness in the central region as dictated by the expected beam size (to minimize scattering) while maintaining strength. Current neutrino factory designs aim at cooling the beam from ~3cm rms radius (~10 cm full size) to ~1cm rms radius over the full cooling channel, and possibly by another factor of 3 (to ~3mm rms or a bit less) for a collider scenario [1].

Safety concerns associated with hydrogen make it necessary to understand the strength of the windows. Historically hydrogen has been used in high energy physics experiments, but the application to muon cooling presents new challenges. The energy deposition in the hydrogen absorber will be much greater and the volume much smaller than in hydrogen bubble chambers and other fixed targets.

Therefore, a clear understanding of the window behavior and strength as a function of applied pressure is required. This has been demonstrated by comparing the deformation of the window as measured via photogrammetry to that predicted by FEA. Photogrammetry is a 3-dimensional coordinate measuring technique that uses photographic images as the fundamental medium for metrology.

Since FEA analyses indicate a strong sensitivity of strain to the window thickness, photogrammetric methods have also been used to measure the thickness of the window, providing more accurate input to the FEA. These measurements have also been compared to the design, providing a method of quality control.

This paper describes:

- Development of photogrammetric methods of stress analysis and quality control
- A brief history of photogrammetry
- Recent developments in photogrammetry
- A description of how photogrammetry works
- Stress analysis, test design, results, and comparison with FEA
- Quality control testing and results
- Measured thickness compared to design
- Evidence that more accurate strain predictions result if these results are fed into the FEA
- Plans and improvements for future windows

## 3. BRIEF HISTORY OF PHOTOGRAMMETRY

The earliest roots of photogrammetry can be traced to Renaissance painters, particularly Leonardo da Vinci, who studied the principles involved in the geometric analysis of pictures in the late 1400s. The next significant development was projective geometry, which forms the mathematical basis of photogrammetry. Notables include Desargues, Pascal, and Lambert, from the mid-1600s to mid-1700s. The actual practice of photogrammetry could not occur until Daguerre's invention of photography in 1839. A year after Daguerre's invention, using kites and balloons for taking aerial photographs, Colonel Aime Laussedat of the French Army directed the first experiments using photogrammetry for topographic mapping. Laussedat is considered the "father of photogrammetry". The invention of the airplane in 1903 facilitated the emergence of modern aerial photogrammetry. Advancements in instrumentation and techniques continued through the last century [2,3].

## 4. RECENT DEVELOPMENTS IN PHOTOGRAMMETRY

Three pieces of equipment are required for photogrammetry: camera, targets, and software. Many recent developments in each of these areas were used to develop methods of photogrammetric stress analysis and quality control.

### 4.1 Digital imaging

Digital photographs have replaced emulsion-based photographs for all but the most exotic applications. Due to advances in digital image processing software, coded targets, and auto-correlation methods, a large number of photogrammetric measurement tasks can now be fully automated [4]. A camera and software system known as V-STARS (Video Stereo Triangulation and Resection Software), from Geodetic Services, Inc. (GSI) was used. V-STARS uses a high-resolution CCD camera to perform digital photogrammetry.

### 4.2 Self-identifying objects

The use of self-identifying coded targets and an artifact known as an Autobar automate the process (Figure 1). The coded targets contain seven square features that are placed in a two-dimensional array, thereby providing a set of unique patterns that is recognized by the scanning software, along with a central circular feature that represents the target's position. The Autobar, shaped like a cross, has six features arranged in a three-dimensional pattern that is recognized by the scanning software. The Autobar acts as a self-identifying reference frame indicator, so that the initial approximations of the object coordinates can be established. In the final stages of the processing, this initial reference frame is generally transformed to a location that better suits the geometry of the object and its analysis. The origin is transformed from the Autobar to the center of the window so that the window flange defines the XY plane and the convex side is in the +Z direction (Figure. 2)

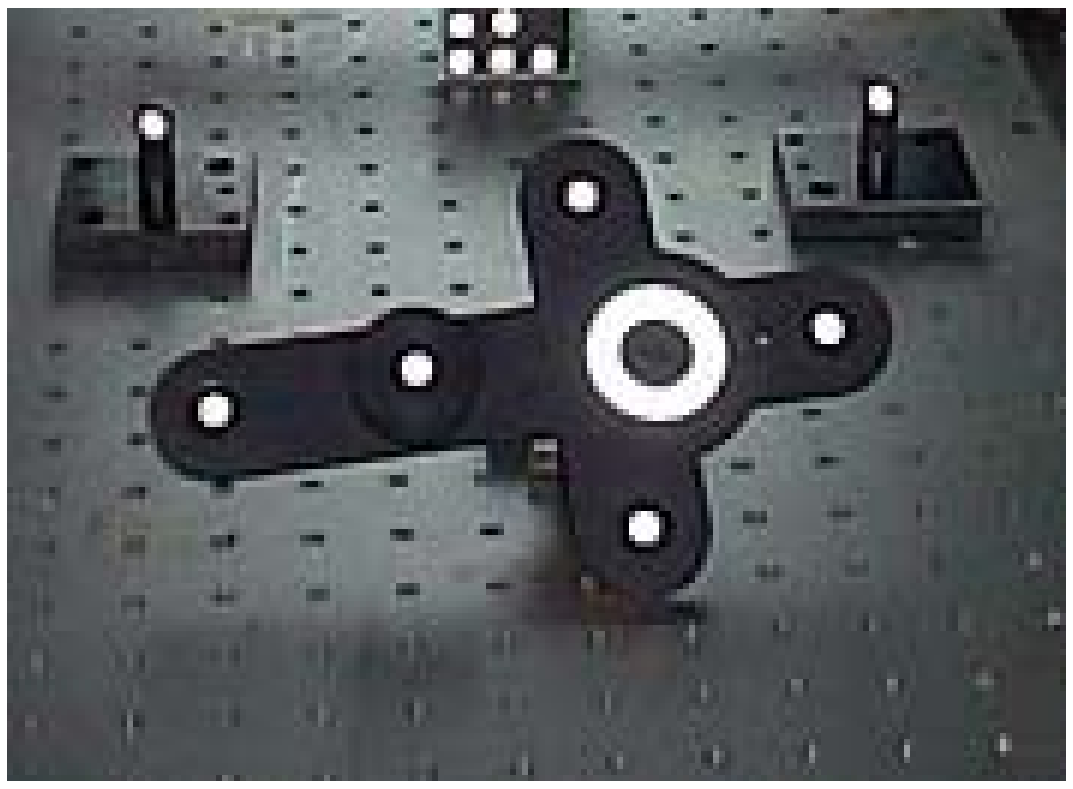

**Figure 1**

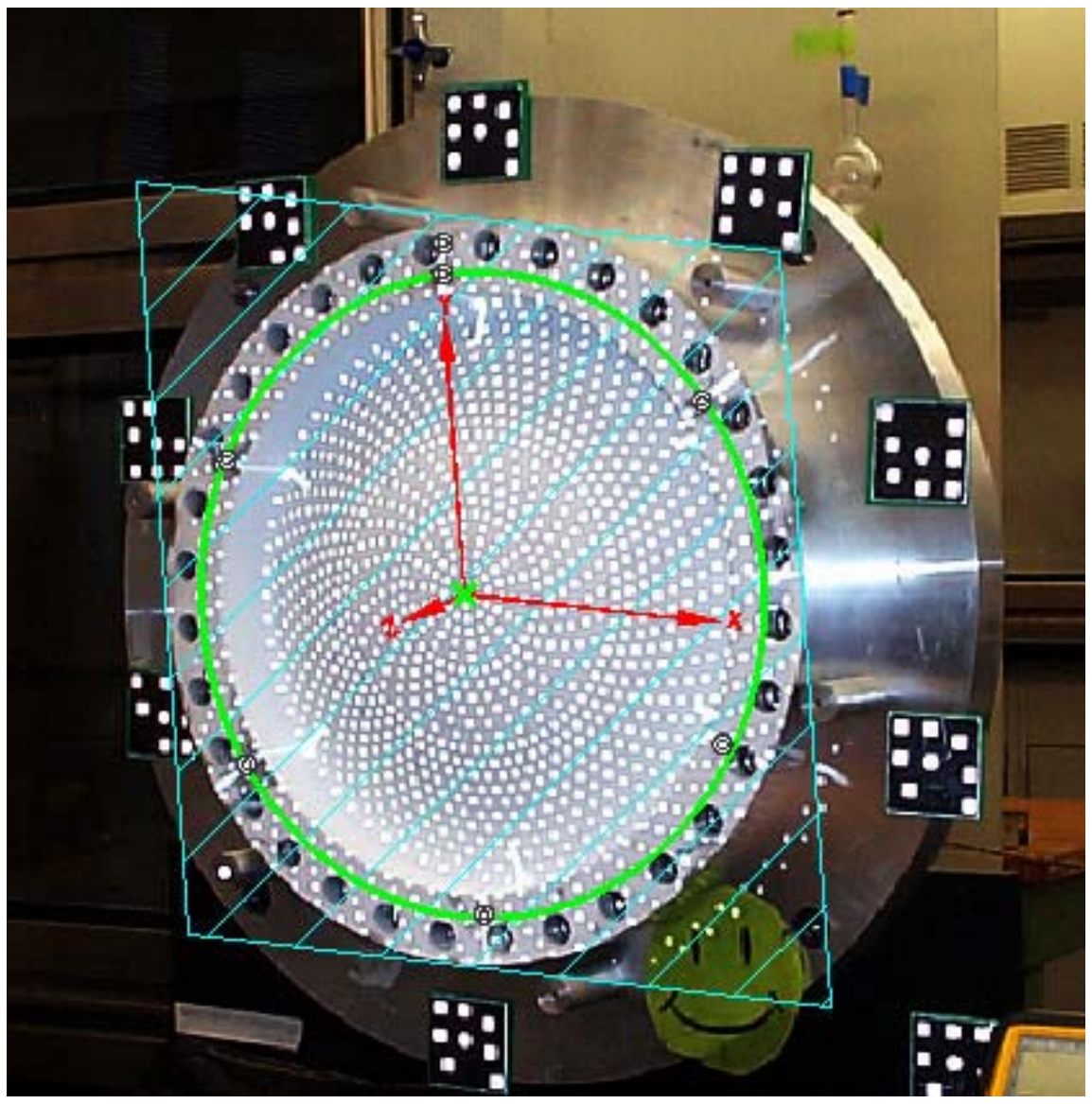

**Figure 2**

### 4.3 Intelligent camera

#### 4.3.1 Hardware

Hardware development has come in the form of the "intelligent camera". The intelligent camera contains an integrated computer and processes the image immediately after it is taken, providing information about the image so that errors, such as poor exposure, insufficient number of coded targets or missing Autobar, disk full, and lens cap on, may be detected and diagnosed during data acquisition [4]. The camera used is a GSI INCA (INtelligent CAmera) with an image size of 18.4x27.6mm, 2044x3072 pixels, fixed focal length of 17mm, and a 56° x 76° field of view. The lens focus is fixed such that the useful depth of field is 0.5m to 30m.

#### 4.3.2 Hardware precision

Measurement precision is a function of the resolution and quality of the camera, the size of the object being measured, the number of photographs taken, the geometric layout of the pictures, and the correctness of the camera calibration. According to GSI, the camera system and software can produce a centroid determination on the order of 1/50 of a pixel by using an intensity-weighted analysis of the target. This value is based on using a single image from a handheld camera, using permanent targets rather than using a projector (see 4.5.1), with a target size of at least 3-pixels by 3-pixels. By examining the details of the design, it is apparent that this is a reasonably conservative assessment of the system resolution. As such, it leaves room for those interested in extracting measurements that are more precise an opportunity to do so. Considering the 3072 pixel (2044) width (height), which measures 27.6mm (18.4), a single pixel measures 9μm square.

At first glance, 9μm would seem to be the lower limit of resolution, however, this is not the case. The following example illustrates this point. Each pixel has an 8-bit range of sensitivity to incident light, which gives an output range of 0 to 255. If a beam of light fills the entire area of a single pixel with sufficient brightness to achieve an output of 255, but without entering into saturation, and all adjacent pixels have an output of 0, the center of the beam can be said to be at the center of the pixel. If the beam is translated such that one of the adjacent pixels has an output of 1 (the other seven adjacent pixels remaining at 0 output), the original pixel will have an output of 254. The translation required to achieve this state is 1/256 of the width of the pixel (9μm /256), or 0.035μm. This value can be considered the fundamental positional resolution of the CCD, at least when intensity weighting is used for position determination. In practice, the value will be subject to ambient light conditions, and near saturation levels, which should be avoided. However, filtering techniques and exposure settings, along with ambient light control, can be used to enhance the signal-to-noise ratio, thereby recovering most of the theoretical resolution.

### 4.3.3 Operational factors

Actual targets are much larger than the minimum 3-pixel by 3-pixel size for several reasons, primary of which is to improve the centroid determination, but also to filter unwanted spurious pixels which may have been illuminated by bright, non-target objects, such as the heads of screws, circuit board pins, etc. In addition to pixel count, shape of the pixel patch is an important filtering criterion in order to reject false targets. The algorithm that GSI uses for centroid determination, as stated above, is based upon the intensity weighting approach but is implemented with some additional, proprietary features.

Without detailing the simulation process used to validate the centroid determination algorithm, it is fair to say that the centroid of a larger patch of pixels is likely to be better determined than a smaller patch of pixels. For our purposes, a target pixel count of approximately 50 was sought. In practice, as it applies to this effort, pixel count usually exceeded 50 and approached 100 from the closer camera stations. What the optimum size of a target is, as it relates to the centroid determination, will be left to the Monte Carlo experts. Certainly, counts above 100 pixels, as it applies to this system, seem to offer little or no advantage.

The functional resolution of the camera, that is the resolution a camera may achieve as a system, can be determined by creating a field of targets, fixing the camera with respect to that field, then taking repeated images of the targets. The X-Y coordinates of the centroid of each target in the image plane should remain constant at the level of functional resolution. As will be detailed later in this paper, this examination is complicated by the inclusion of the projector.

The following items contribute to the precision of a photogrammetric solution:

- Resolution of the camera
- Structural integrity of the camera (whether the CCD moves with respect to the focal point of the lens)
- Dimensional uniformity (does the camera "grow" detrimentally in response to the heat generated by the on-board computer, for example)
- Correctness of the camera calibration
- Other, more subtle, factors which may or may not be known
- Stability of the scene
- Geometry of the camera positions relative to the object
- Number of exposures
- Object-space size

### 4.3.4 Measurement precision

The precision of a piece of work, that is the precision of the system as it is applied to the project at hand, is governed first by the requirements of the end user. In this case, with a predicted maximum deflection of about 2mm, a precision of 10μm in the direction of deflection (+Z direction) would be sufficient, but with the objective of improving positional precision in the X-Y direction by refinement of method.

An examination of the size of the object, the available geometrically suitable camera locations, and the practical working distance from the camera to the object, dictated that the distance from camera to the object would be in the range of 0.7m to 1.2m or, nominally 1.0m. Considering the 17mm lens of the camera, a magnification ratio of approximately 60, ((1.0 +0.017)/0.017) is realized. Multiplying this value by the fundamental resolution (0.035μm) yields a lower limit of working resolution of 2μm.

## 4.4 Refinement of method

Initial efforts indicated that surface deformation was being measured with a precision of 8 to 15μm, which were determined from between 50–60 images (precision of the components of the target position are determined as part of the V-STARS bundle adjustment.) While the results were at a nearly acceptable level, the transverse components were larger than expected, as high as 0.2mm, and quite irregular. While examining the causes of this issue, a number of opportunities were discovered that would allow for an improvement in overall precision. The culprit in these early efforts was the mounting method for the projector on the optical bench.

By careful refinement of the procedures used, along with some of the system components, a precision in the range of 3.5 to 5μm has been achieved. This was done primarily by stabilizing the components of the system, examining the medium term (several hours) stability of the camera and projector, and developing optimum geometry for the camera locations.

The stability test was done by taking a series of close range (0.5m) images of the target pattern (about 5mm diameter), as projected onto a coated aluminum sheet, over a period of several hours in parts of two days, at ten-minute intervals. The aluminum sheet was affixed to a granite table, as was the projector, while the camera was mounted on its standard tripod. A cable release was used to isolate the camera from the observer. The first image was used as a reference, with each successive image being compared to it.

Approximately 2,000 targets were analyzed for each image. While a shift was noted between the two days of about 1.5μm in X and Y in the image plane, the internal consistency of each several-hour session was in the range of 0.1-0.4μm in X and Y. This procedure examines the effective stability of the camera and the projector as a single system. As the numbers suggest, our stabilization effort was successful.

As for geometry, the goal was to take as few images as possible during each pressure epoch. A rule of thumb offered by GSI states that a minimum of twelve well distributed camera positions would be sufficient to ensure optimal results. Since the most useful indicator of stress from a metrology point of view is displacement in the Z direction, a bias in the camera positions was sought in order to enhance the parallactic angle. This was done by exaggerating the air-base, the separation between camera stations in the X and Y directions.

### 4.5.1 Projected targets

One limiting factor that has affected the acceptance of photogrammetry in general is the need for point-of-interest targeting [4]. If the object is fragile, it may be undesirable to touch the object to apply standard retro-reflective targets, and if the area or desired density of target coverage is large, it may be very time consuming to adequately cover the area. Hence, the focus of recent developments has been to eliminate the necessity to physically target the measured object. Since the absorber window is very thin (~330um at the center) and very complete coverage of the area was desired, a non-contact targeting method was attractive.

A new, non-contact targeting technique that employs a high-power stroboscopic projector, GSI's PRO-SPOT, to project a pattern of dots (targets) onto a surface, was selected (Figures 2 and 3). The dots are of high contrast and quality, mimicking conventional retro-reflective targets, but have no inherent thickness. This feature is especially attractive when the technique is applied to measuring the window dimensions for quality control. The results can be directly compared to design without any compensation for target thickness. Although a thin layer of a flat-white coating was required to achieve high contrast on the reflective aluminum surface, it was shown that the measured window dimensions were not affected. In the testing of Window 4, an effort was made to measure the thickness of the material by measuring both faces of the window. First, measurements were made with the standard coating procedure, followed then by a measurement set using an extremely light 'dusting' of the coating material. The comparison of the two sets of measurements yielded a difference of 3.6μm for the flange thickness, as determined by calculating the best-fit planes from approximately 200 targets on each surface.

While the flange measurements where successful, the thinner coating caused contrast deficiencies on the spherical surface because of excessive surface reflection from the aluminum. This resulted in poor centroid determination and, consequently, poorer error estimates than the prior set. An attempt was made to recoat the surface as in the first case, but poor technique and lack of patience caused excessive running of the coating material. This resulted in an increase in apparent thickness of about 30μm. While the deformation measurements are not sensitive to the coating thickness, it is clear that proper technique in applying the coating is required for thickness measurements.

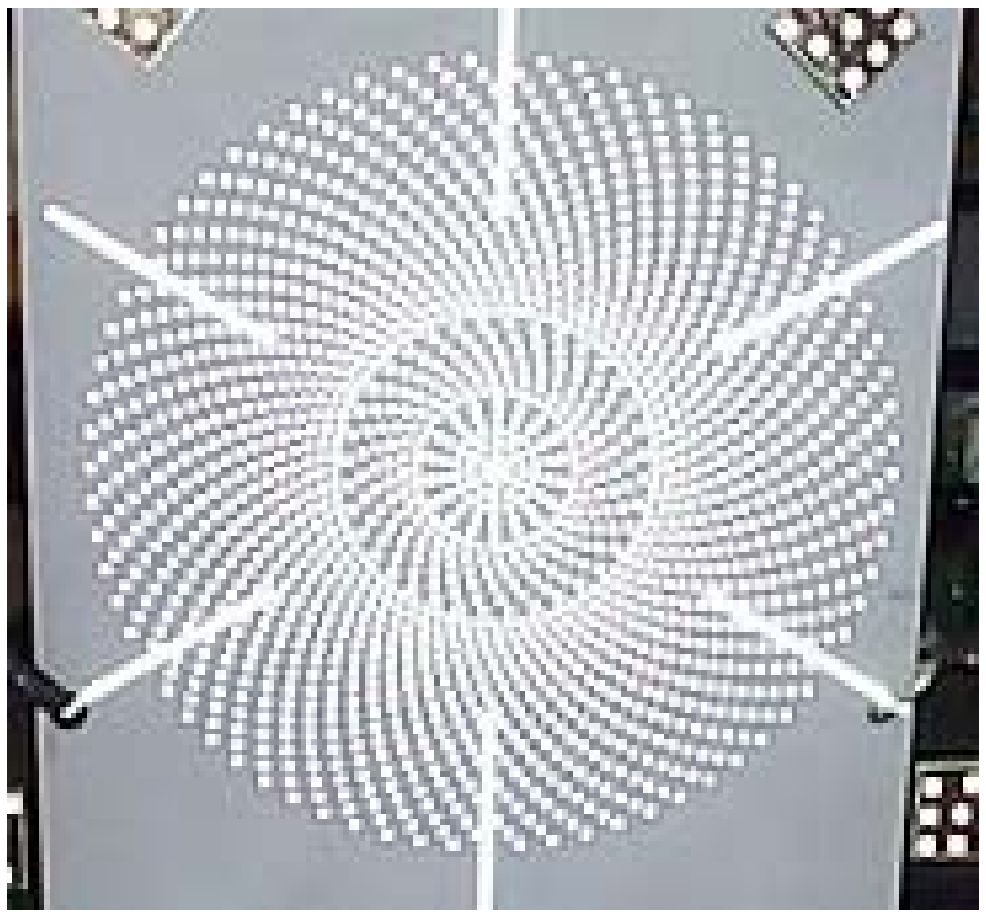

**Figure 3**

The principle of the projection system is much like an ordinary slide projector. A strobed light source, triggered by the camera flash (Figure 4) illuminates a target slide. This illuminated pattern passes through a series of lenses that magnify the slide and project it onto the object. The actual construction of the projector is complicated by the need to accurately control the whole process. By far, the greatest concern is the stability of the dot pattern. Instability of the pattern is tantamount to moving the object during the measurement [4].

As was mentioned above, there was evidence of projector instability during tests of Window 1 and Window 2. If the projection system is stable, the coordinates of the targets that are orthogonal (in the X-Y plane) to the projector should remain constant throughout the test. The instability was attributed to motion of the projector and not motion of the windows, by noting that only the coordinates of the projected targets changed, while the retro-reflective targets attached to the windows remained constant. Increased stability was achieved through the design of a more stable mount for the projector (Figure 5) and by operating the camera from a tripod with a remote trigger.

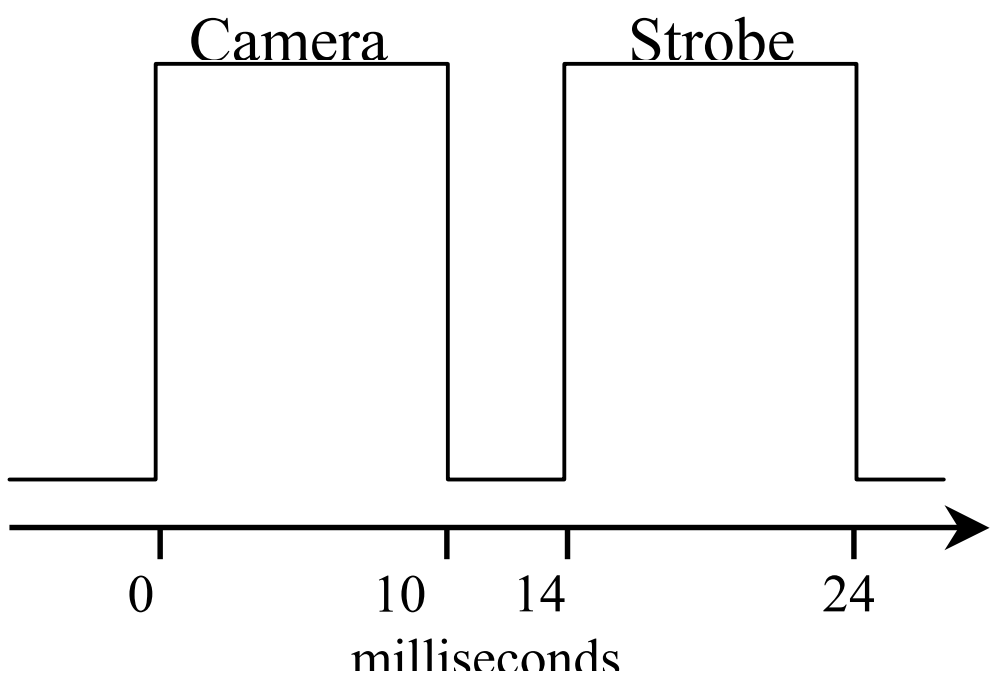

**Figure 4**

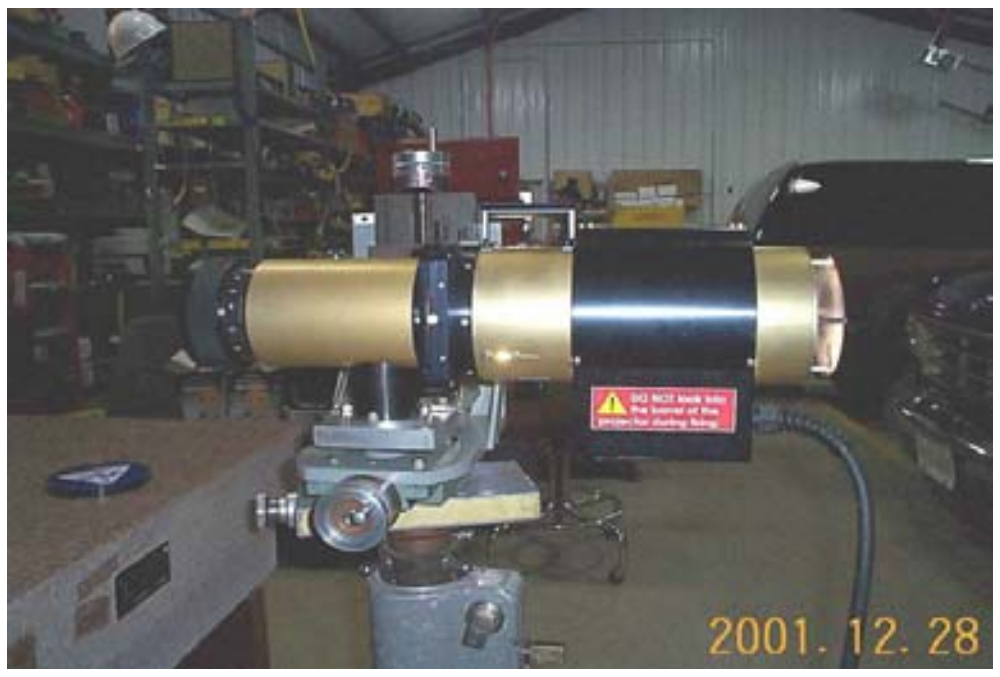

**Figure 5**

### 4.5.2 Target pattern

The target pattern is created by an array of holes in the slide. The slide provided with the projector was used for the tests of Window 1 and Window 2. This slide has a rectangular array of 5600 dots, but only 200 of the dots were projected on the window surface at the desired dot size. To sample a larger fraction of the window surface and to better suit the circular geometry of the window, a custom slide with a radial pattern was designed which increased coverage by an order of magnitude (Figure 6). The holes in the slide are 0.2mm in diameter on 0.4mm centers. With a magnification of about 15, this resulted in 3mm-diameter targets on the window.

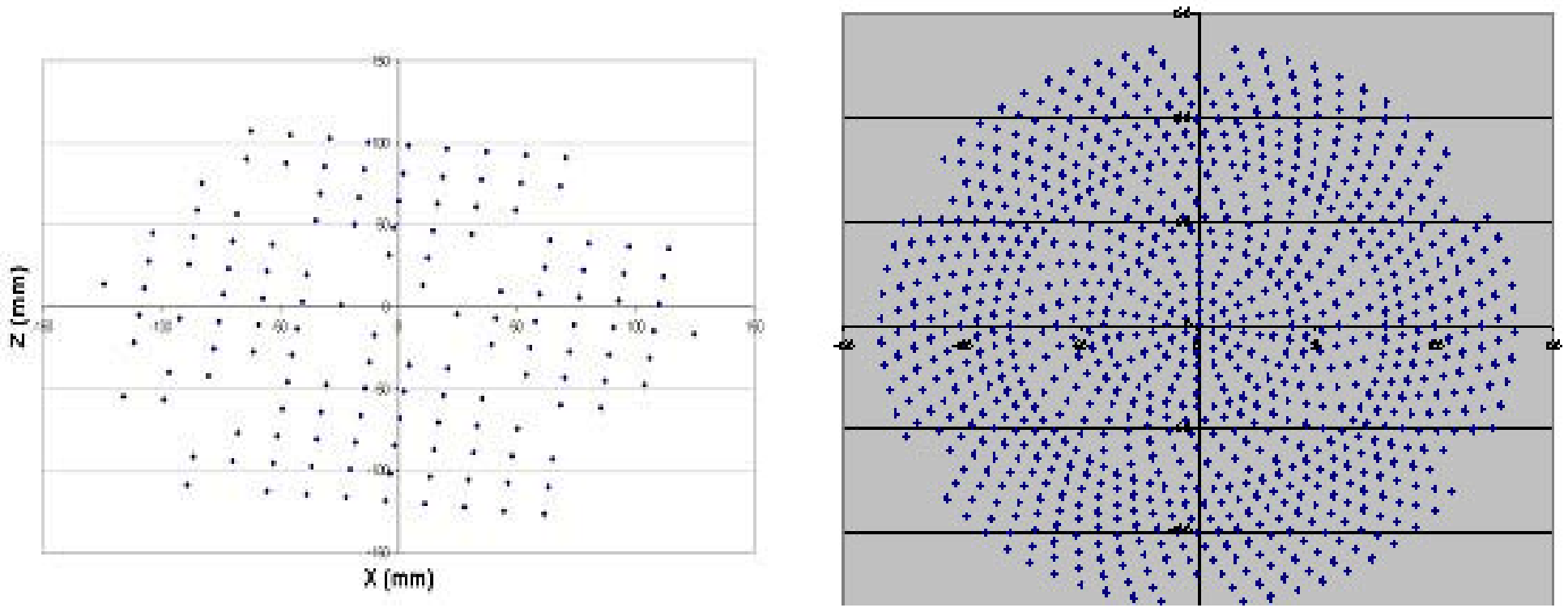

**Figure 6**

The radial pattern proved very effective in tests of Window 3 and Window 4, providing more data near the center of the window. This is the location of prime interest since it is the thinnest, weakest region of the window.

## 5. THE PHOTOGRAMMETRIC PROCESS

### 5.1.1 Photogrammetric principles

Photogrammetry uses the principle of parallax to determine the coordinates of the object of interest. Parallax is the apparent shift in an object's position when viewed alternately from different vantage points. Parallax is the primary basis upon which our eyes gauge distance within our surroundings. Distance and parallax go hand in hand. A more distant object has a smaller perceived parallax.

### 5.1.2 Astronomic analogue

Astronomers use parallax to determine stellar distances by measuring the apparent shift of a relatively nearby star against the background of identifiable, more distant stars due to the motion of the earth around the sun (Figure 7). The angular distance between the background stars (which exhibit negligible parallax) provide a scale to measure the shift. Half the baseline is 1AU. This angle is twice the parallax angle, *p*, and forms a right triangle with the baseline and line of sight to the star. Using the small angle approximation for tangent, the distance can be calculated: d=1AU/*p*.

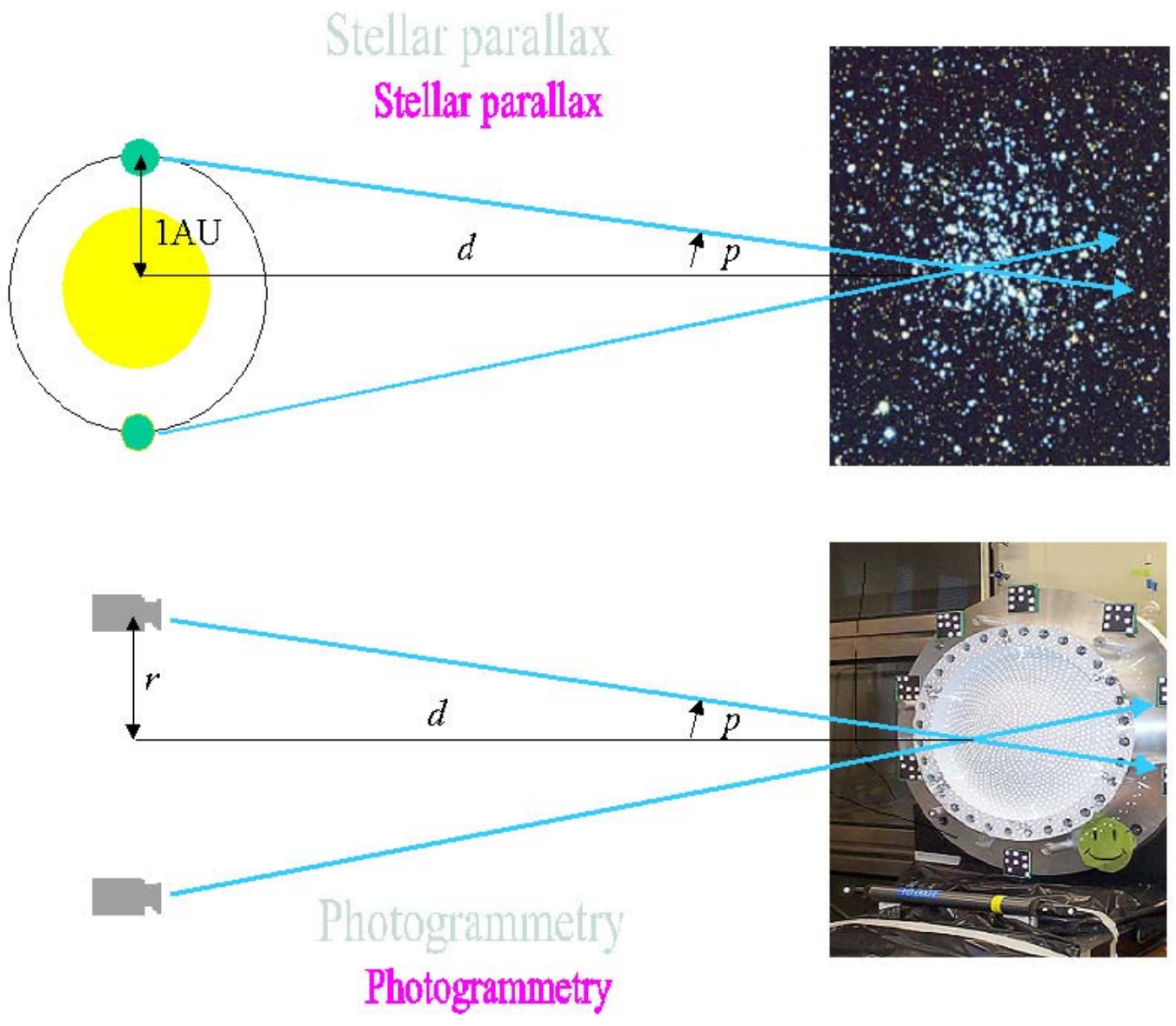

**Figure 7**

Similarly, photogrammetric measurements are initially dimensionless. Dimensions are established by use of one or more scale artifacts, generally of Invar or carbon fiber construction (Figure 8).

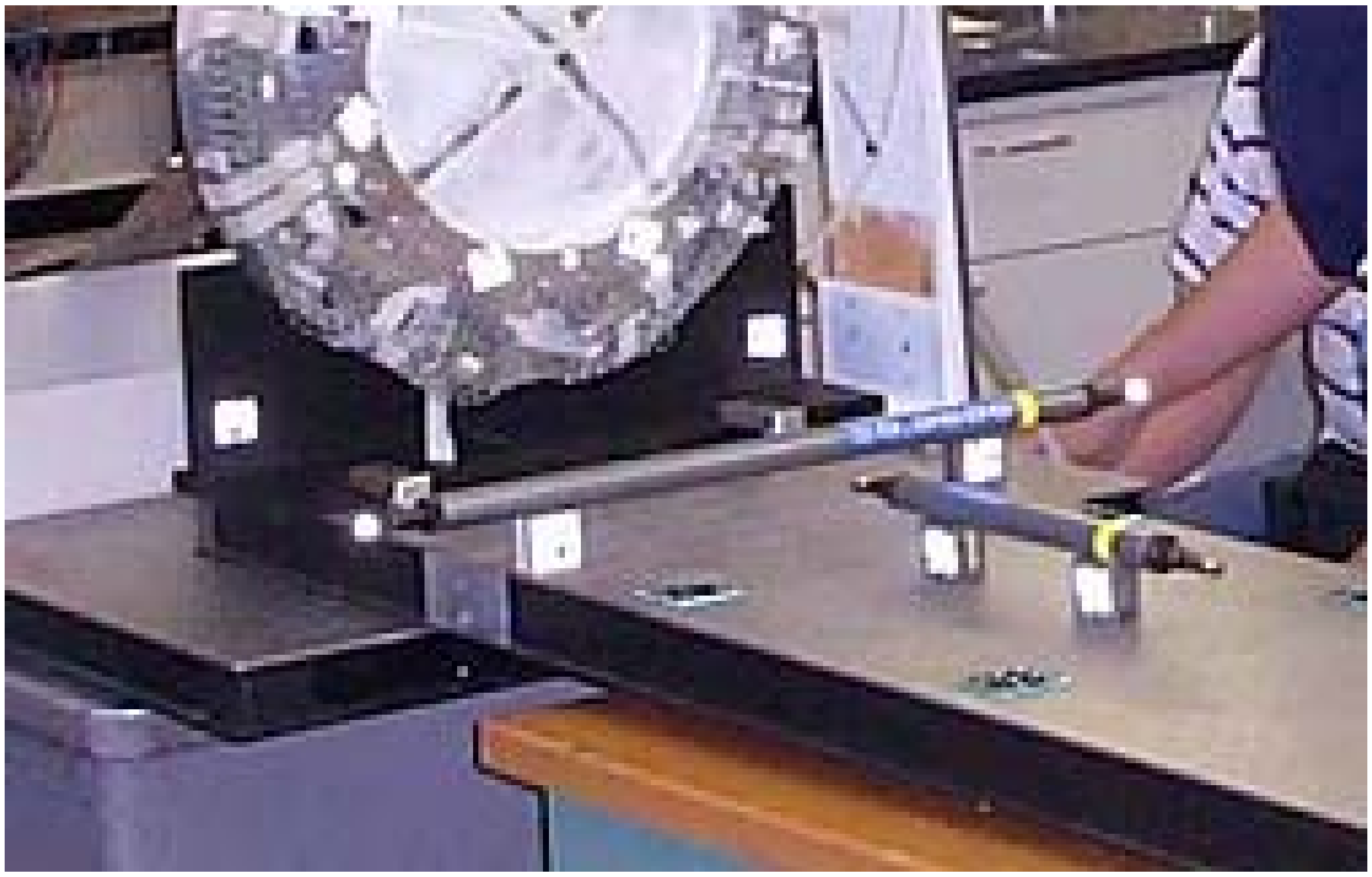

**Figure 8**

Notice that the length of the baseline used in the stellar parallax measurement, the distance between the earth and sun, must be known from an independent measurement. Historically, this was measured by observing the parallax of planets. The modern method is to measure the distance to Venus using radar techniques (Appendix B).

Different from stellar parallax calculations for astronomy, where the baseline must be determined by independent means (Figure 9), in convergent terrestrial photogrammetry the solution is based on the principle of ray-intersection. In this type of photogrammetry, the cameras do not need to be, and are generally not, located at predetermined positions. The photogrammetric solution is accomplished by the use of collinearity equation solution [2], [3], where a ray is a line that connects a point in object space and that object's image in the CCD (image space). In a perfectly calibrated camera, all rays for a single image pass through the focal point of the camera. If several of the objects have previously established coordinates, the position of the focal point in object space (X, Y, Z) and the orientation of the image (roll, pitch, yaw, or $\Omega$, $\phi$, $\kappa$) can be determined. This process is known as *resection*. Once these parameters have been determined from images taken from several different camera locations, the positions of the unknown targets can be calculated by intersecting rays from these additional targets (Figure.10). This process is known as *triangulation*. The process is repetitive, in that as unknown targets become known, more images may be resected and more targets may be triangulated.  Resection and triangulation are repeated until all the coordinates of all objects in all images have been determined.

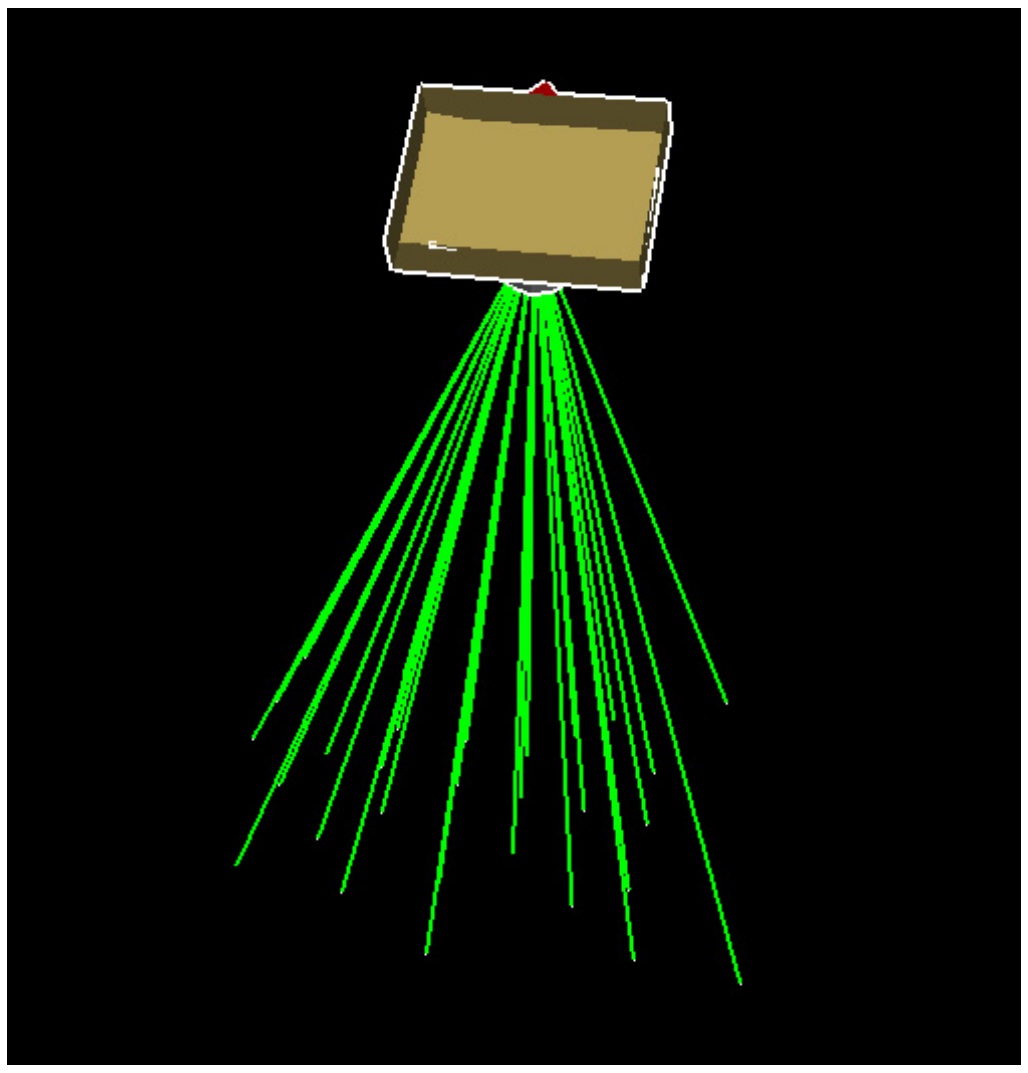

**Figure 9**

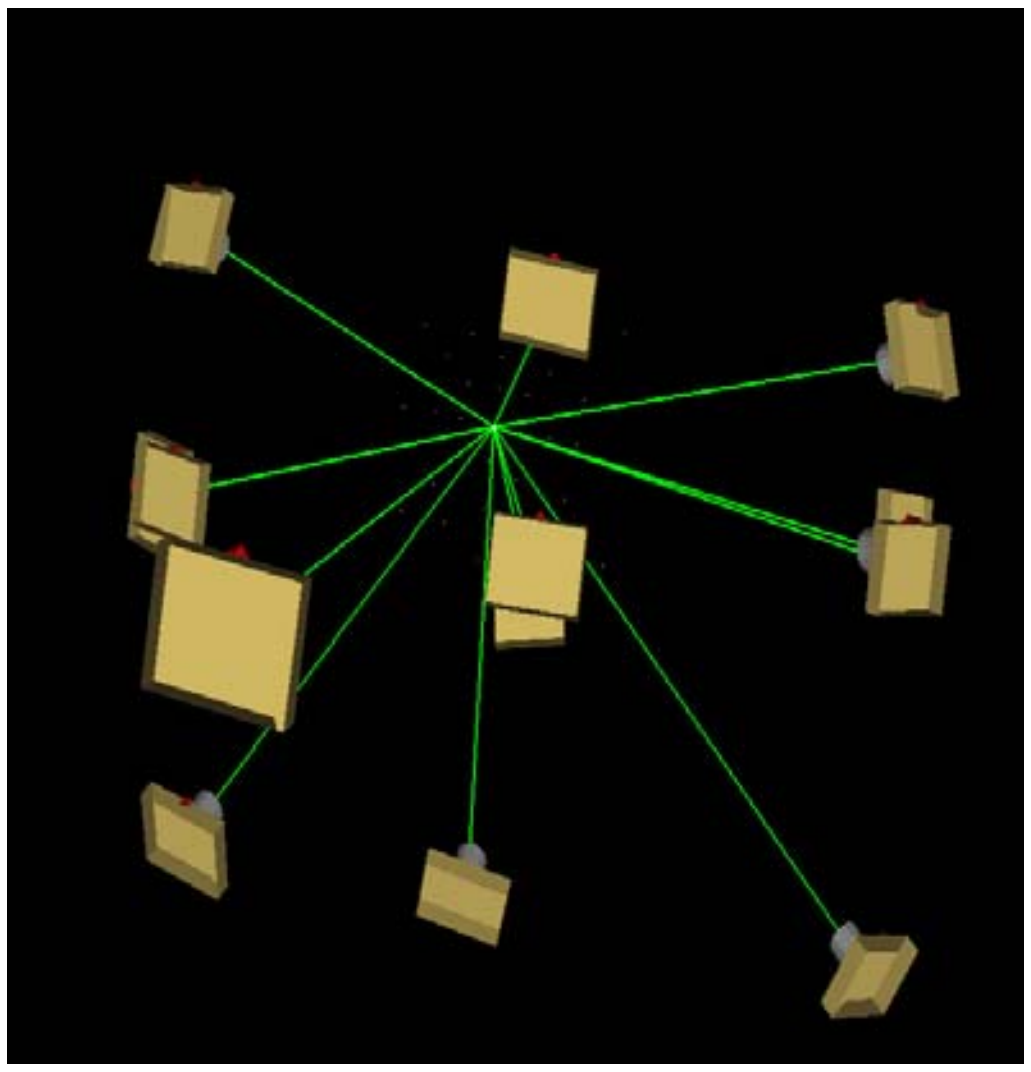

**Figure 10**

### 5.2 System options – single camera vs. multiple camera

As when performing stellar parallax measurements, the photographs do not need to be taken simultaneously.  In modern photogrammetry there are two general options for acquiring images: 1) multiple cameras set at various positions around the area of interest, with the images taken simultaneously; or 2) a single camera which is moved to various positions around the area of interest, with the images taken sequentially.  The choice is whether the volatility of the scene is so great that the scene is likely to change significantly in a short period, or whether the highest possible precision is required and the scene can be expected to be stable over a reasonably short period.  Two issues come into play: 1) camera calibration and 2) the number of rays available to make the coordinate calculation.

In a multi-camera system, the calibration of each camera is based upon a previous analysis, usually by the manufacturer, which may have been done under different conditions than those present during the subsequent measurement.  In a single-camera system, a process known as self-calibration is part of the measurement solution.  Self-calibration is a computational process that refines the camera's calibration as required by ambient conditions, such as camera temperature. By taking many images with the camera in the normal orientation, as well as rolled around its principal axis (rotated 90 degrees), sufficient information is available to determine deviations from the camera's nominal calibration.

The single-camera system was used, because the highest possible precision was required and the scene could be expected to remain stable for the time necessary to acquire twelve images. Of these twelve images, two were taken with the camera rolled 90 degrees (Figure 11). The time required to acquire the twelve images was typically between sixty and seventy seconds.

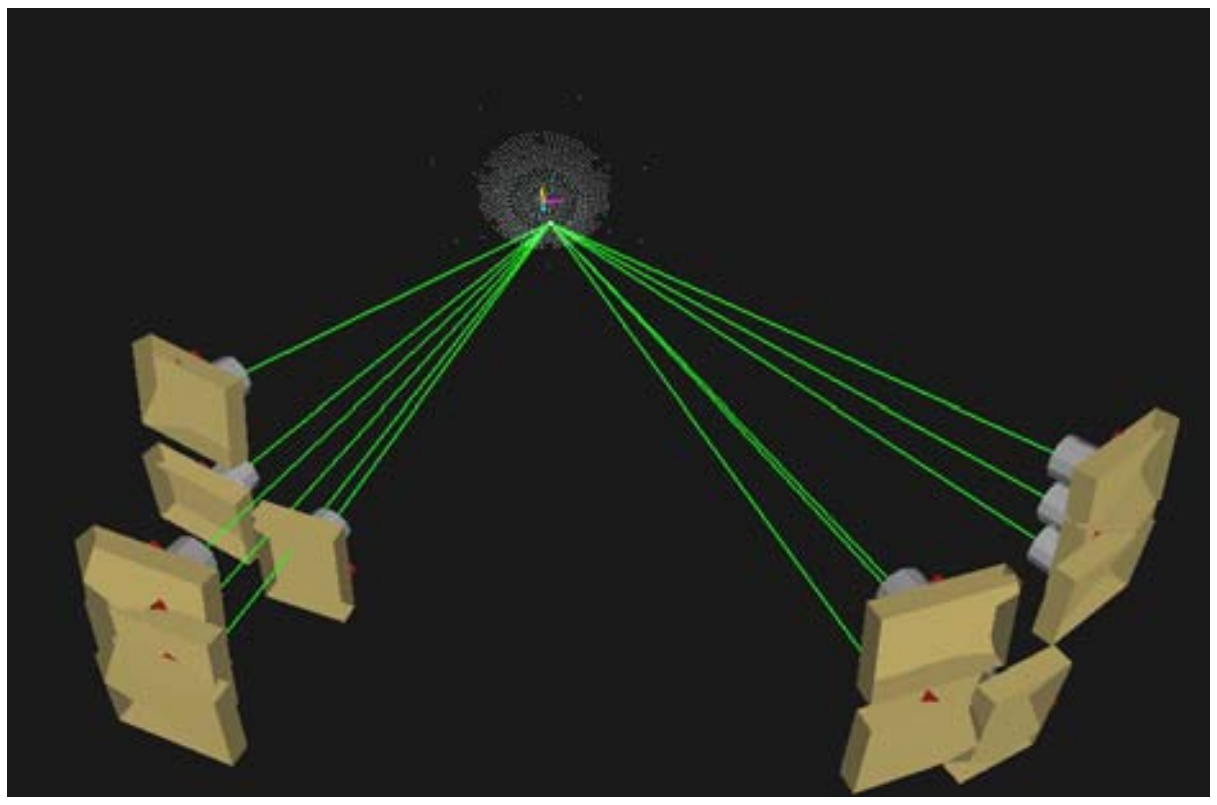

**Figure 11**

### 5.3 Photogrammetric data reduction

Astronomers do not rely on only two snapshots of a star at the extreme ends of the Earth's orbit to determine its distance. As many observations of a target star as possible are recorded. The angular shift is incredibly tiny, and it is easily lost in the noise. Similar to the calibration required in photogrammetry, the data must be corrected for effects of earth's atmosphere and for imperfections in the telescope optics. Astronomers want to be convinced that parallax, and only parallax, can explain the shift observed [5].

In the same spirit, more than two images are used in photogrammetry. Twelve-ray intersections are typical. Since as few as two rays can determine a 3D coordinate, the position of the object is generally over-determined. A least squares analysis, known as a bundle adjustment, is performed to determine the most probable coordinates for each object, the error estimates, the camera's calibration, and estimates of precision for the entire process. The intersection of the twelve rays results in an ellipsoid of uncertainty. The V-STARS Bundle summary reports the X, Y, Z components of this ellipsoid for each measured point (Figure 12).

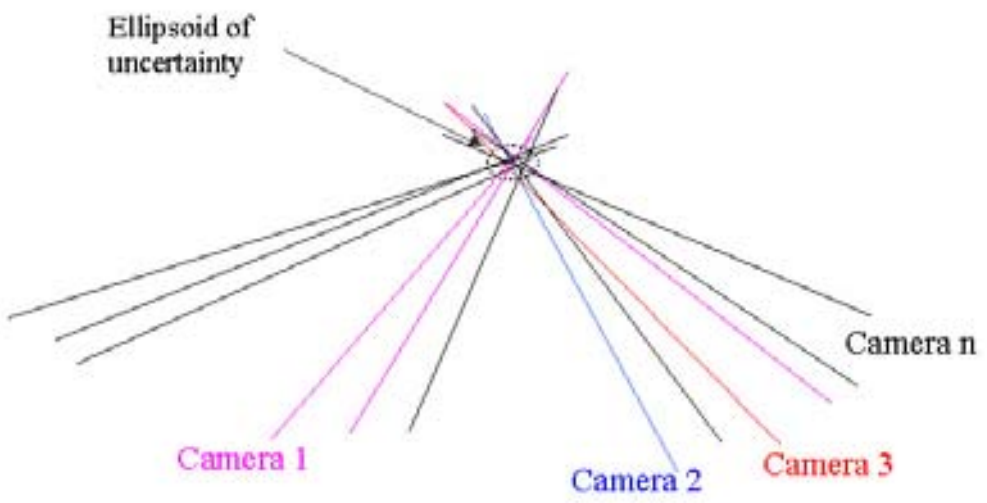

**Figure 12**

## 6. STRESS ANALYSIS VIA PHOTOGRAMMETRY

After initially testing both strain gages and photogrammetry, photogrammetry became the method of choice to measure the behavior of the window under varying amounts of stress. Photogrammetry was preferable to the more standard strain gages for measuring strain for two reasons: 1) Using non-contact targeting for photogrammetry, there is no danger that the tests will damage or alter the fragile windows; and 2) the coverage is much more complete.

### 6.1 Initial test design (Window 1 and Window 2)

The first two windows tested were instrumented with strain gages and measured via photogrammetry. The measured quantity in each test is different. Strain gages measure the tangential elongation of the material, $\Delta L/L$, while photogrammetry measures the deflection in the Z direction (the targets are projected orthogonal to the window flange). Unlike strain gages that are glued to the surface, the targets do not exhibit any motion in the X-Y direction in response to stress. Therefore, it is not possible to directly compare the strain gage and photogrammetry data.

However, FEA can simulate both strain and deflection as a function of stress. If both types of measurements agree with the FEA, it may be concluded that both are viable means to measure window behavior.

Several strain gages were applied along four radial lines separated by 90 degrees (Figure 13). Each gage was read out via a Wheatstone bridge with a temperature-compensating gage in the parallel leg.

The tests were performed at room temperature with stress applied by pressurized water.

The point of yield was consistent between evidence from both strain gages and photogrammetry.

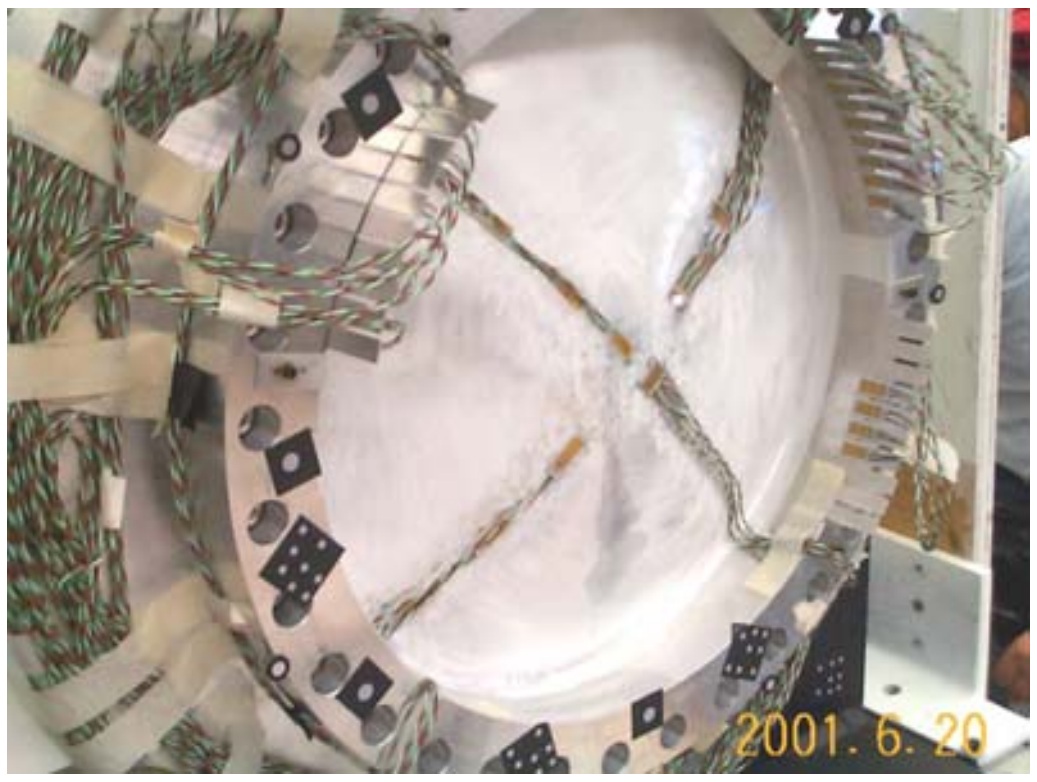

**Figure 13**

## 6.2 Data from new test setup (Window 3 and Window 4)

As mentioned above, the data from Window 1 and Window 2 inspired improvements in the setup stability and photogrammetric technique. Hence, the bulk of the data discussed in this paper will be from Window 3 and Window 4.

The pressure versus yield characteristic, for several radii (in the X-Y plane), can be seen in (Figure 14). The data provide a clear picture of yield at lower pressures for smaller radii, where the window was much thinner, while no yield was seen at larger radii, where the window was thicker.

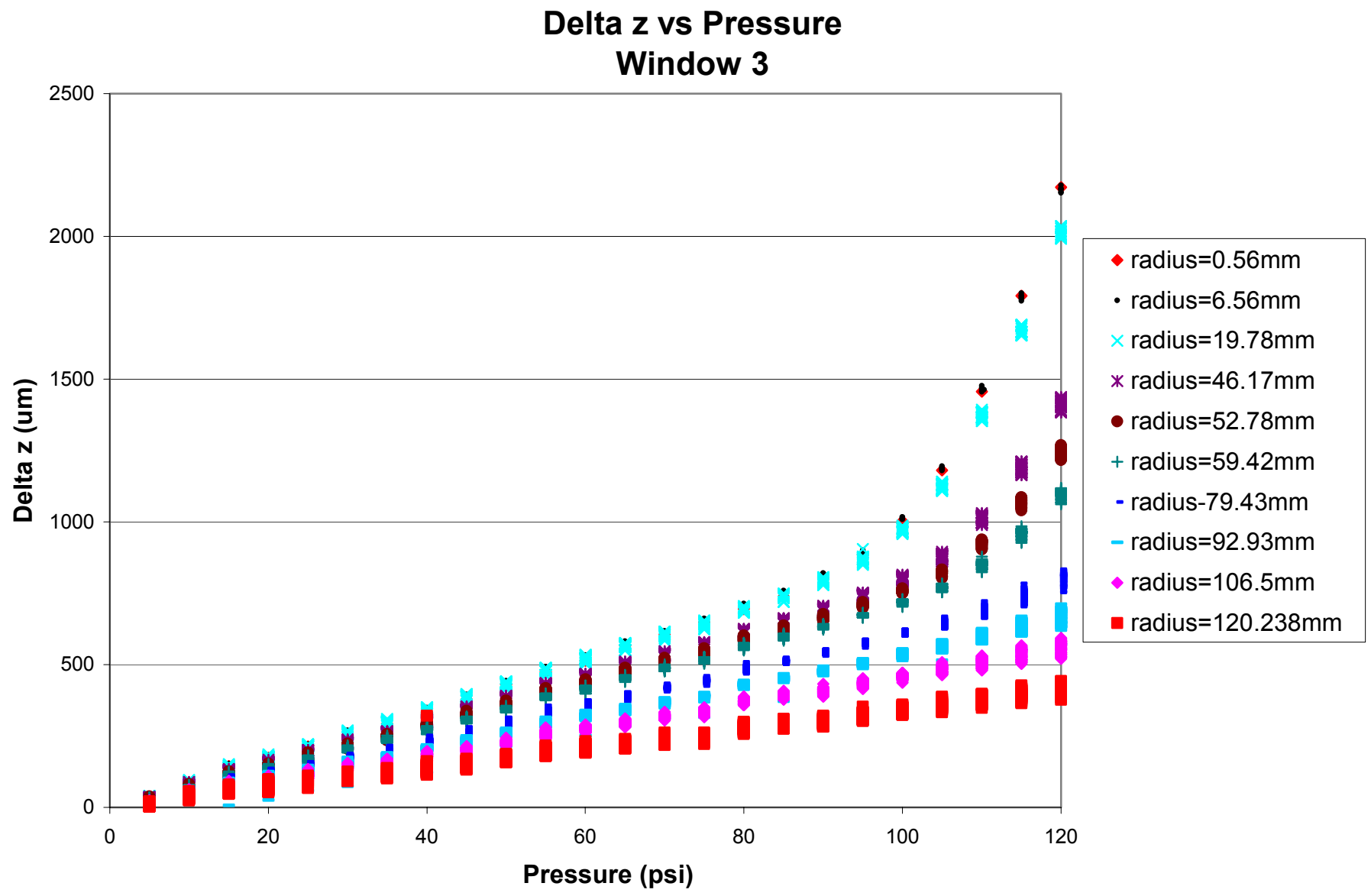

**Figure 14**

## 6.3 Comparison to FEA

The FEA includes prediction of the behavior of the window in the linear region to the point of ultimate yield. The deflection from zero applied pressure of the windows described by the coordinates derived from photogrammetry, may be compared to the deflection predicted by FEA for each test pressure. Representative results for Window 3 plotted as "deflection (delta Z) vs. pressure", for increasing radii as shown in (Figures 15a-15f).

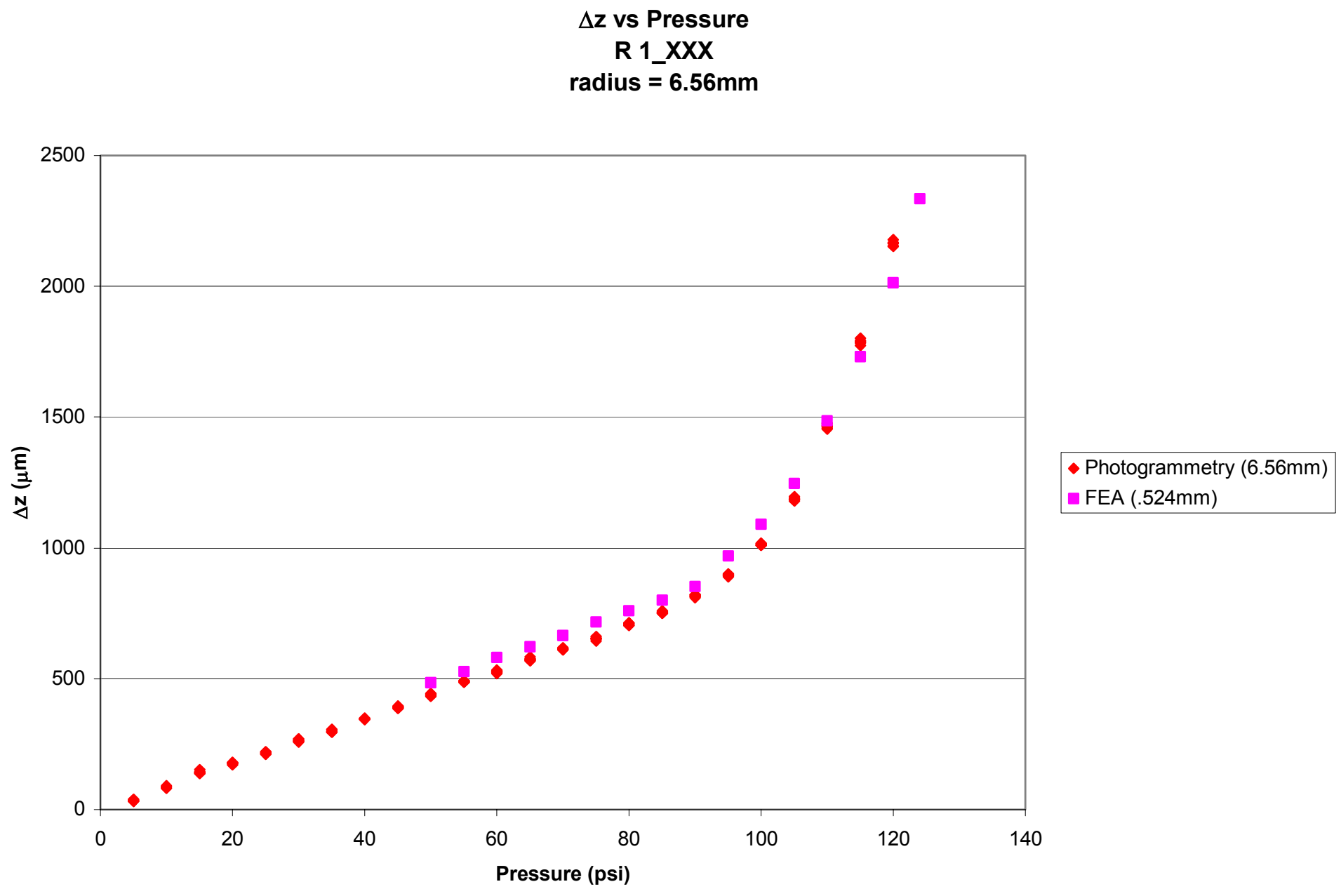

# Figure 15a

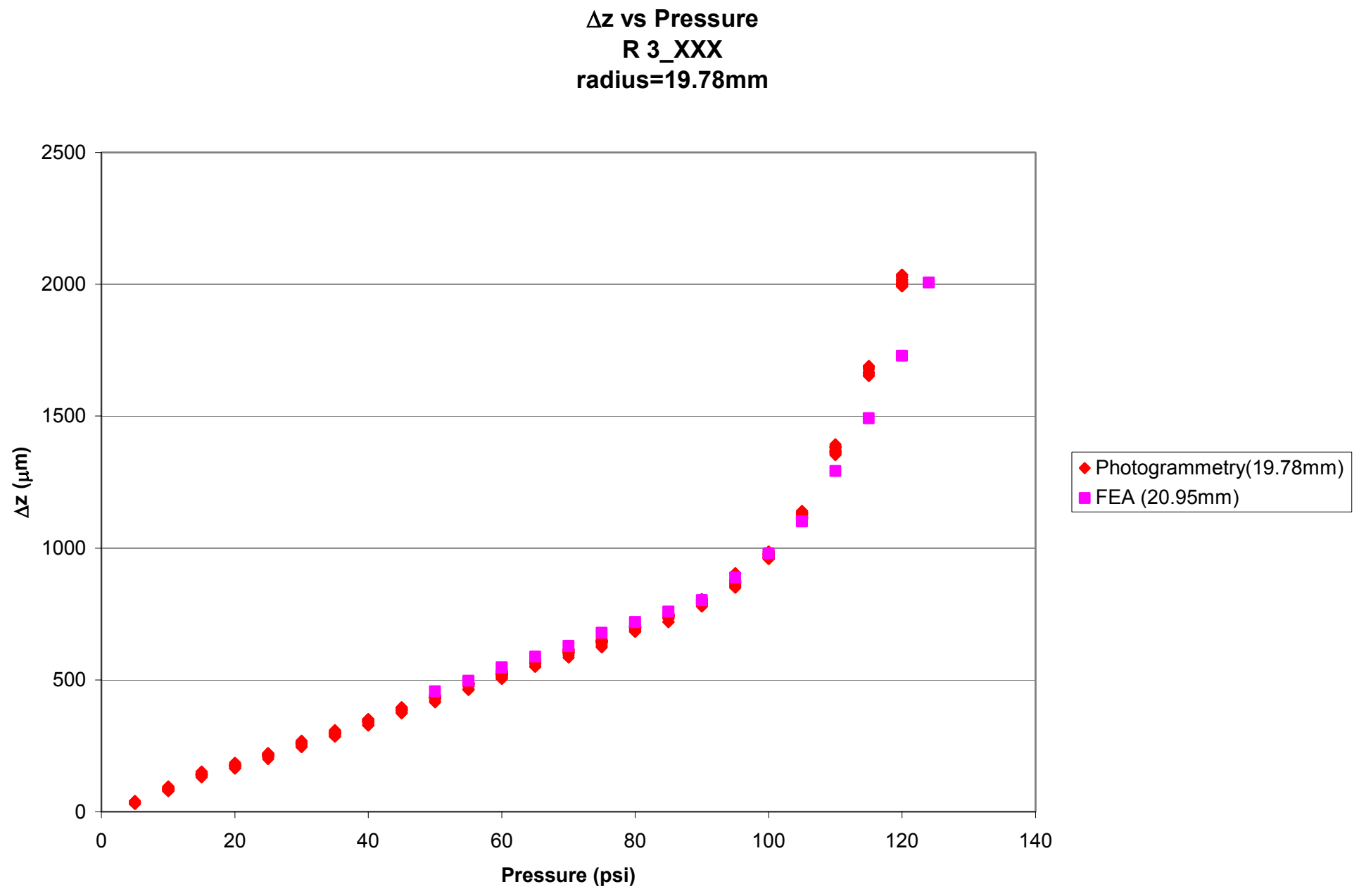

# Figure 15b

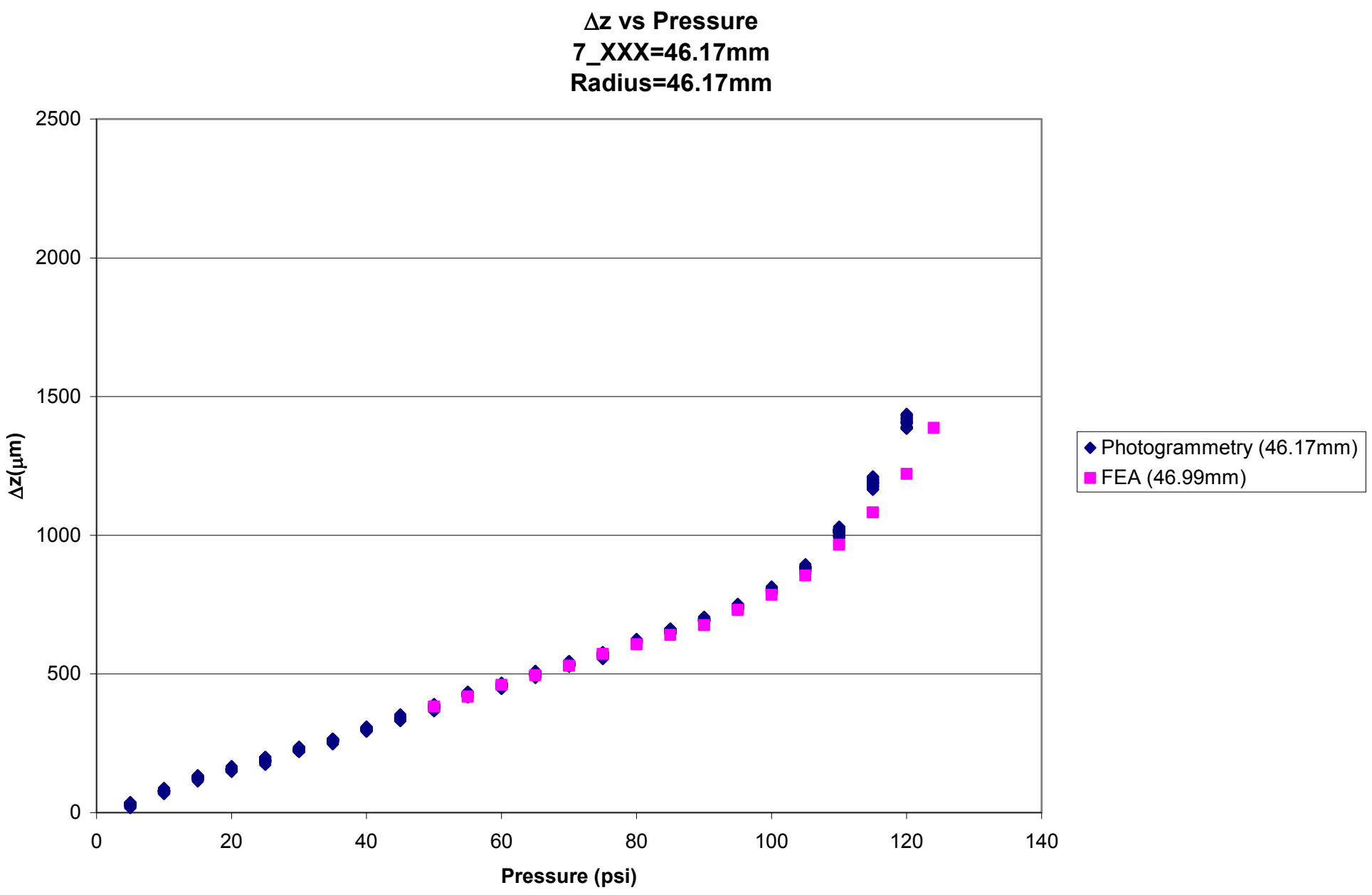

**Figure 15c**

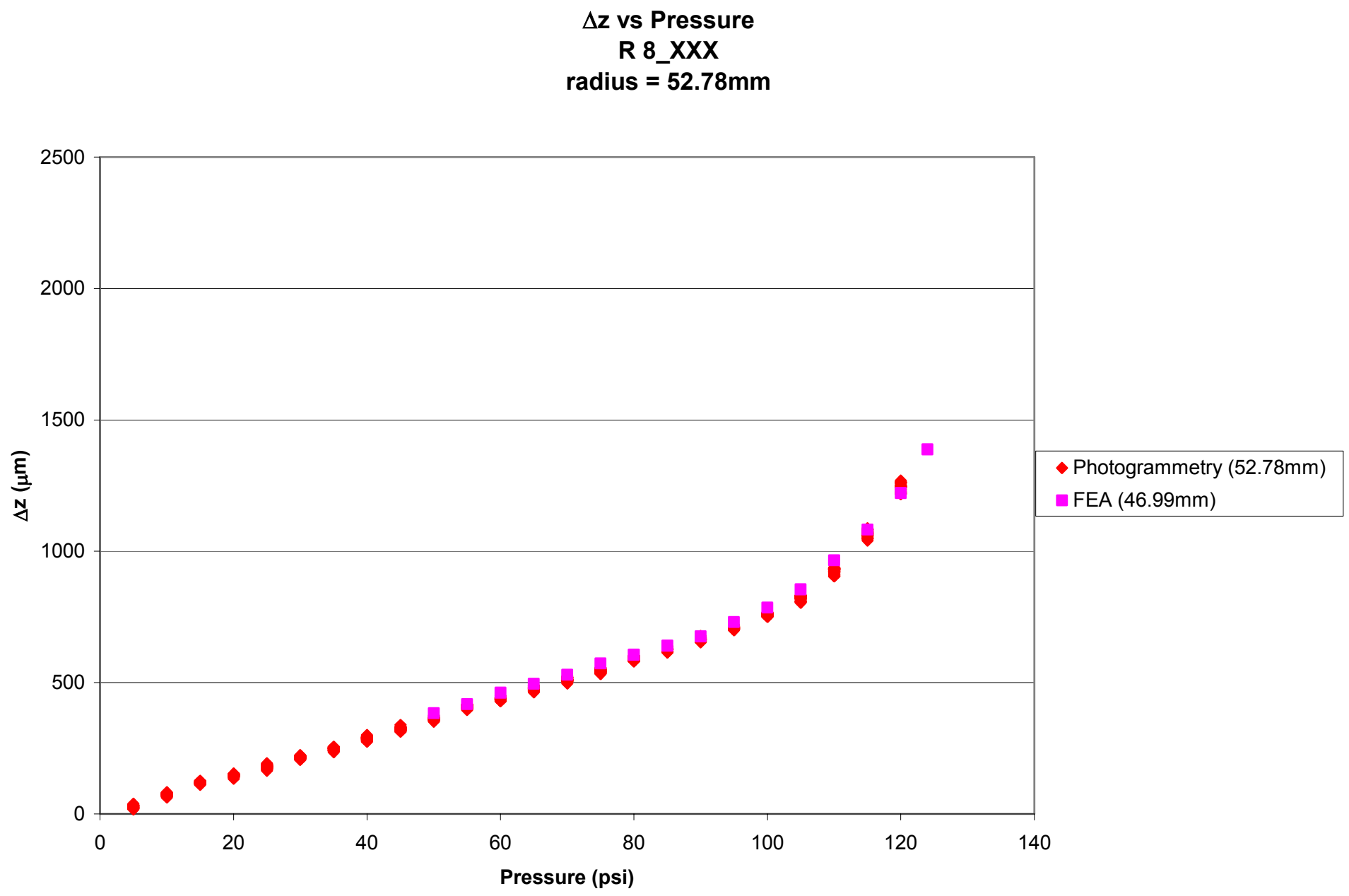

**Figure 15d**

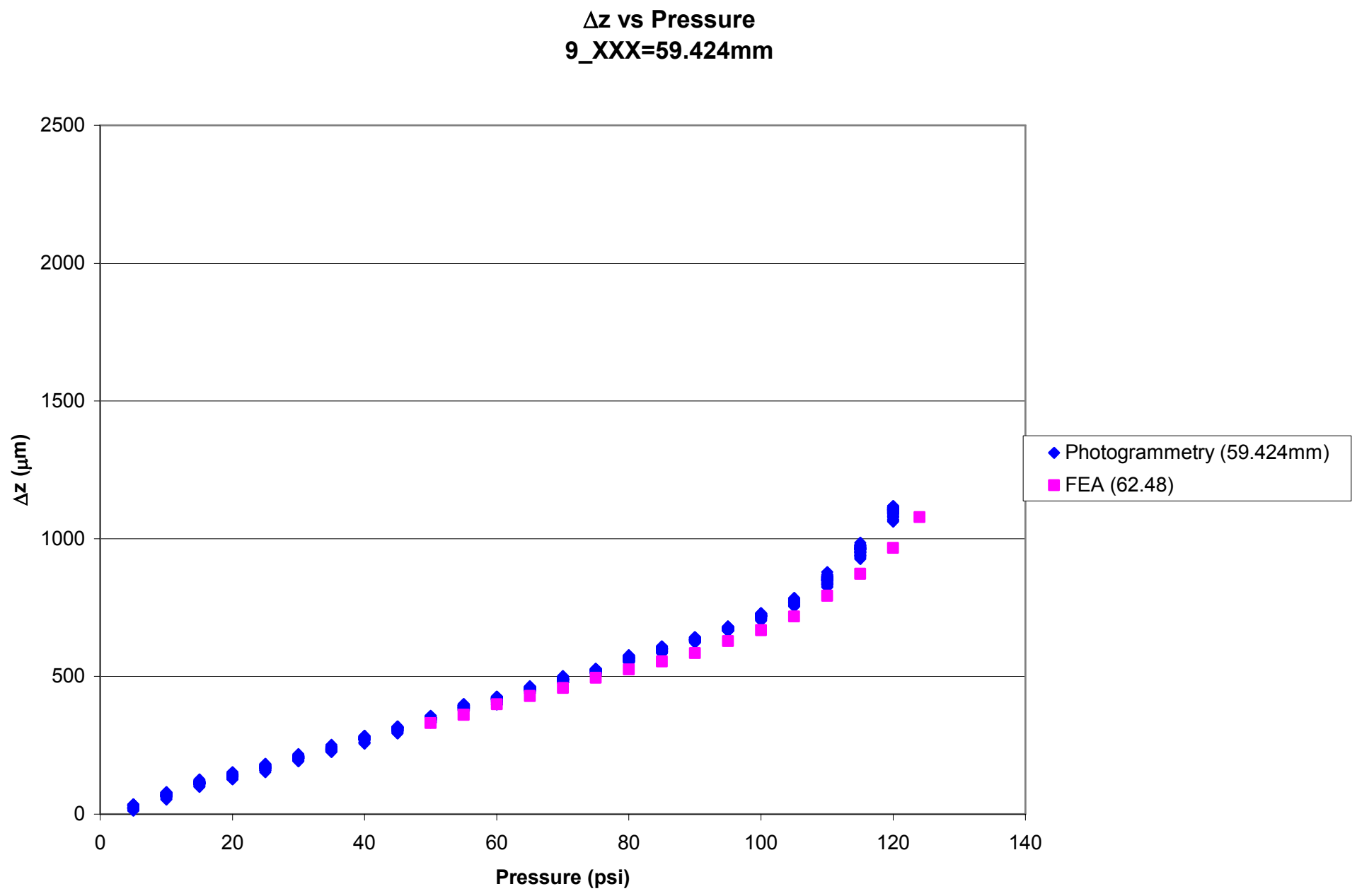

**Figure 15e**

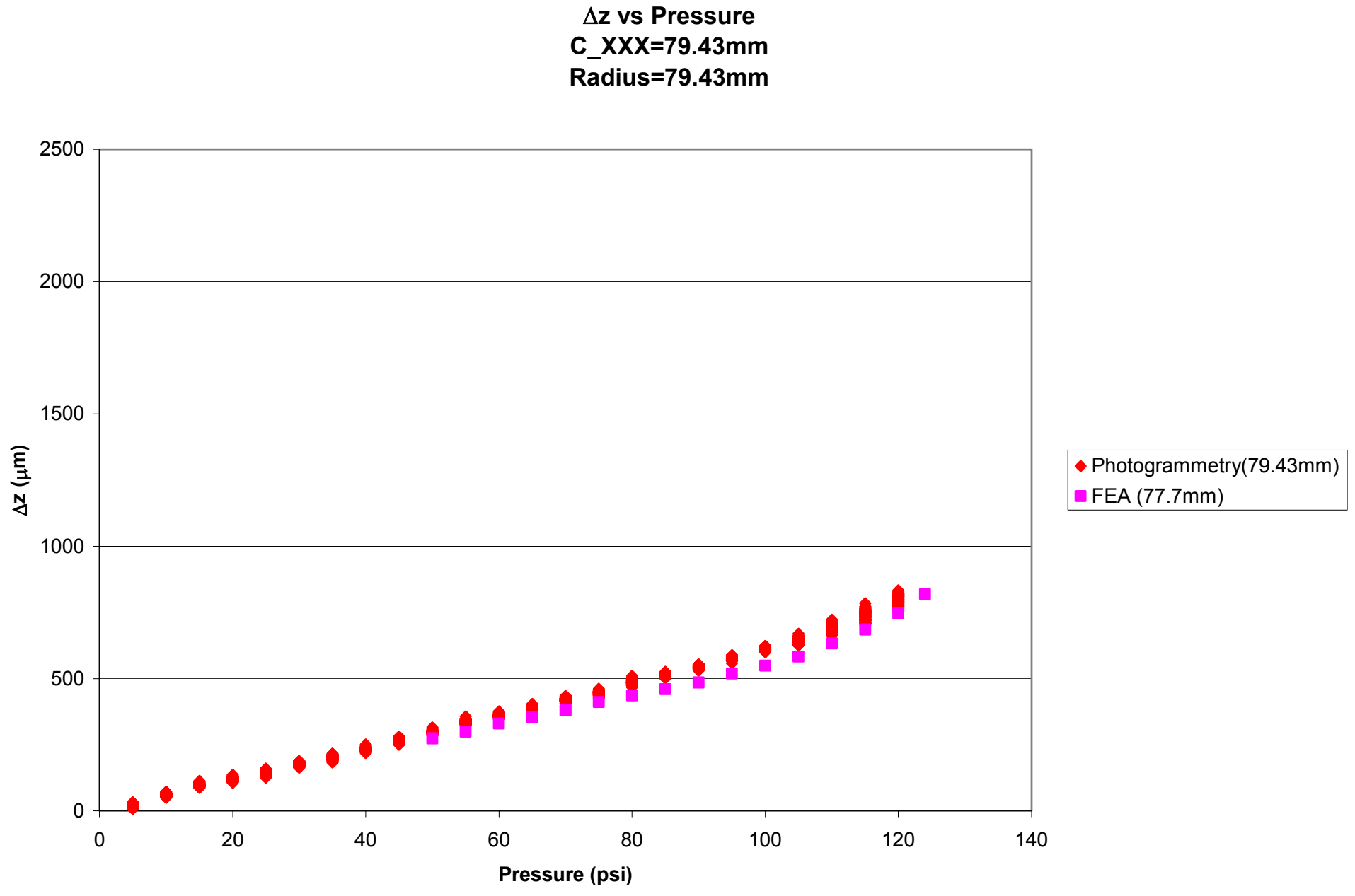

**Figure 15f**

Window 4 was tested at liquid nitrogen (LN) temperatures. The dewar used did not permit photogrammetric measurements to be made. Hence, photogrammetric measurements were made at room temperature up to 50psi (well below yield stress), to ensure that material characteristics were not changed prior to the LN test. These measurements were compared to the data from Window 3. The agreement between Window 3 data and Window 4 data (Figure 16a-16c) indicates the consistency in manufacturing and the viability of using photogrammetry for material control for future production runs. While measurement of the window shape (see 6.5) provides information about the geometric conformity to design, it would not reveal information about any flaws in the material. Measuring the window's response to stress is required for this test.

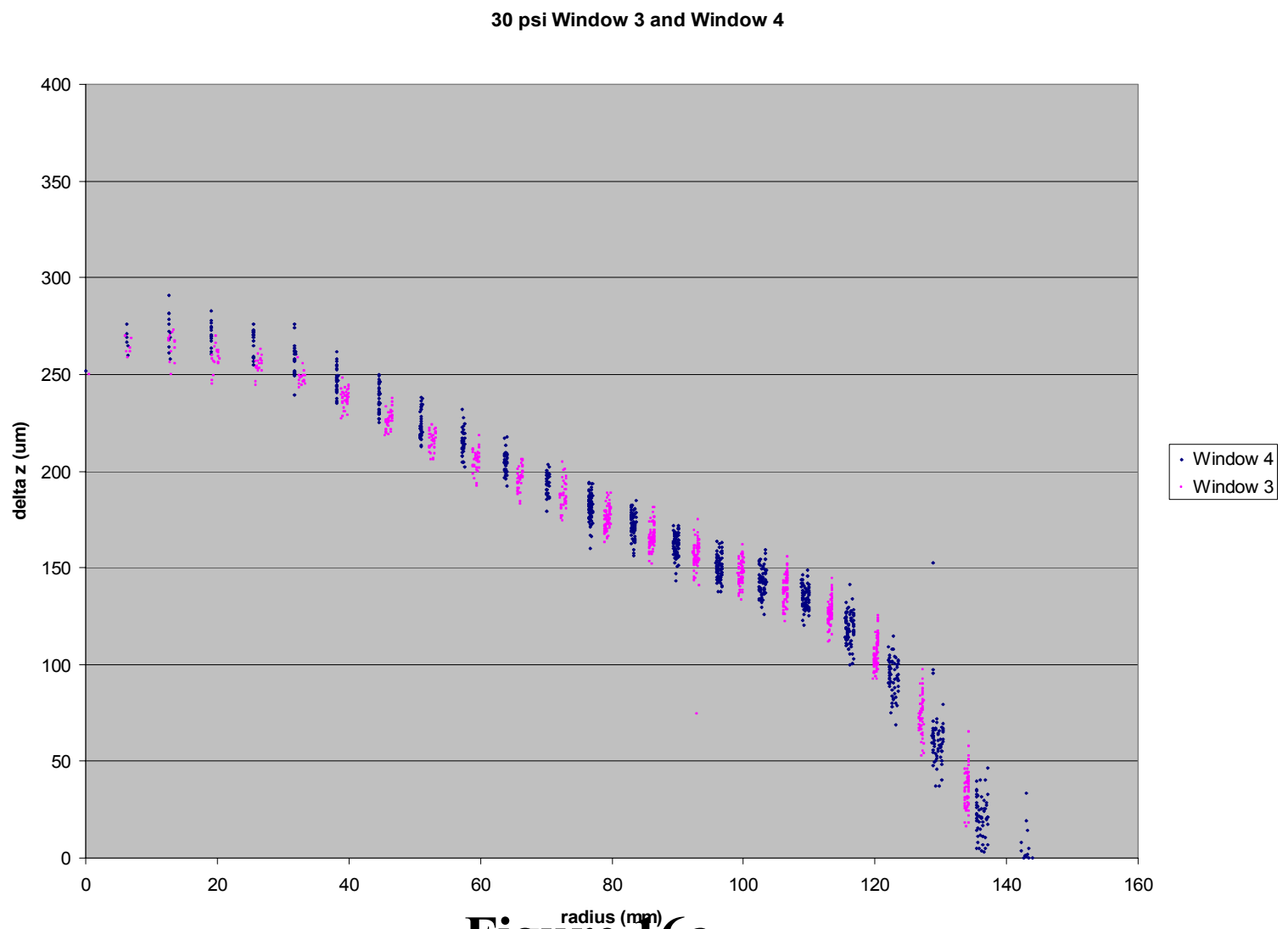

**Figure 16a**

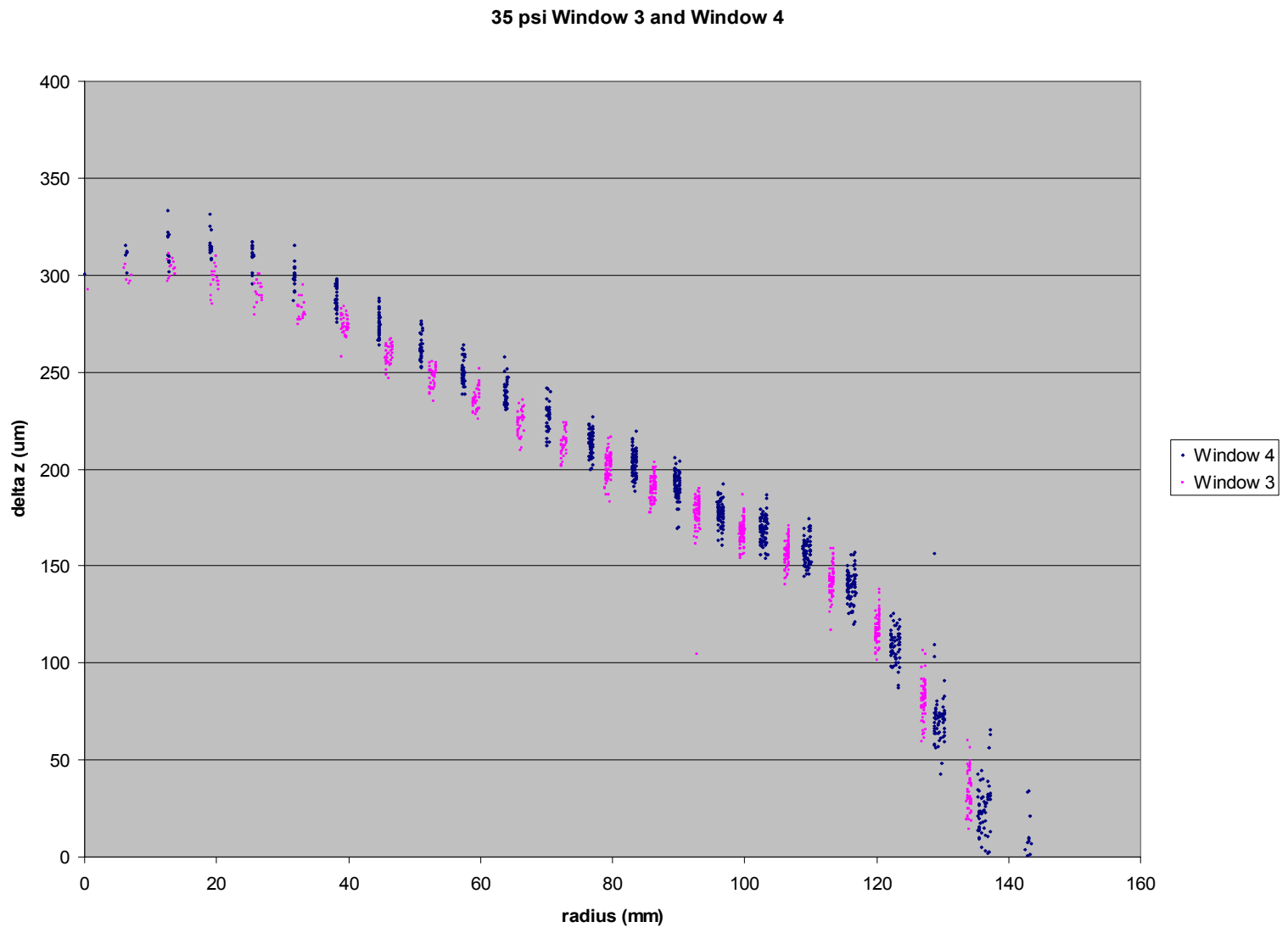

**Figure 16b**

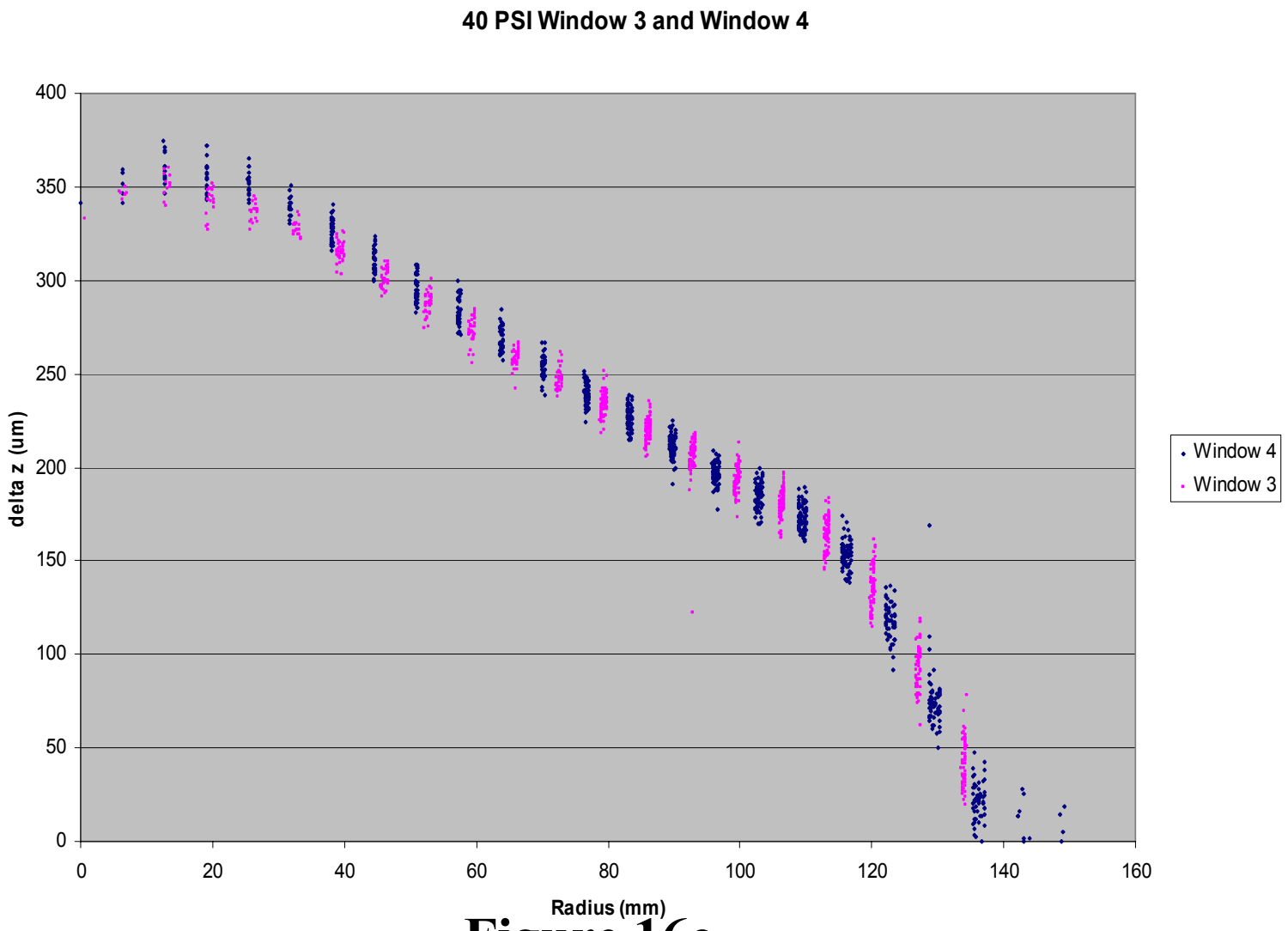

**Figure 16c**

To comply with safety requirements, it was necessary to determine the pressure at which the window burst. All four windows tested burst very near the pressure predicted by the FEA (Table 1).

| | Test temperature | FEA results | | Test results | |
|---|---|---|---|---|---|
| | | Minimum window thickness (mm) | Rupture pressure (psi) | Window thickness from CMM (mm) | Measured rupture pressure (psi) |
| | | | | | |
| Window 1 | 293K | 0.11 | 48 | 0.11 | 42 |
| Window 2 | 293K | 0.33 | 117 | 0.33 | 119 |
| Window 3 | 293K | 0.345 | 123 | 0.345 | 120 |
| Window 4 | 80K | 0.33 / .348 | 156 / 162 | 0.33* / 0.36 | 152 |

**Table 1. FEA results and test results**
*** thickness from photogrammetry**

### 6.4 Benefits of extensive area coverage

The very extensive area coverage provided by photogrammetry permitted observation of a harmonic effect (Figure. 17). This effect is under investigation. Several explanations for the effect have been offered, such: as an offset target pattern; window shape eccentricity; actual static harmonic oscillations in the window; and, perhaps, a superposition of several effects. These effects would not have been visible in the strain gage data.

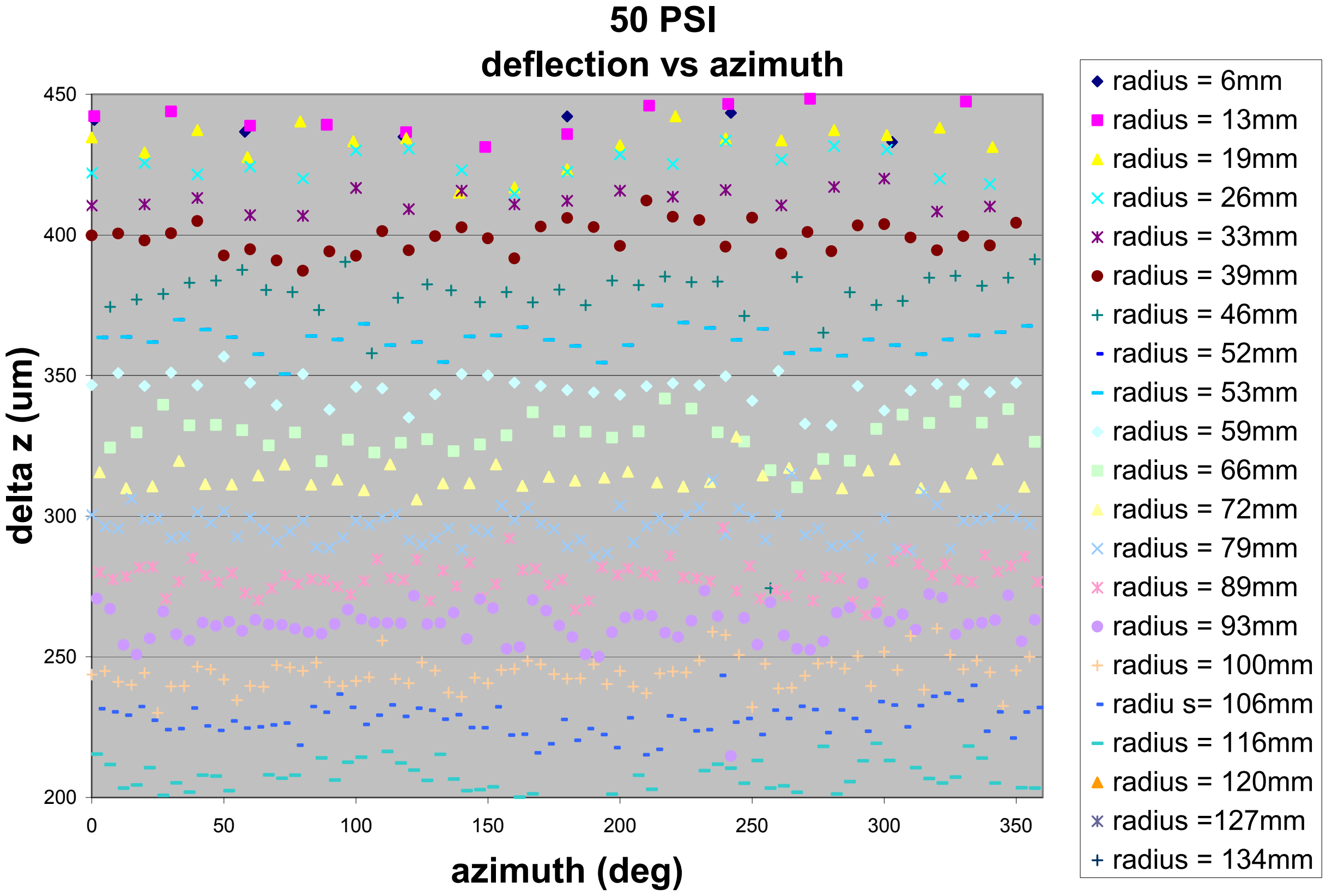

**Figure 17**

### 6.5 Material control (compare to design)

Photogrammetry was also applied in comparing the window dimensions to design. This complements the initial method of material control provided by measurements made with a coordinate measuring machine (CMM). Photogrammetry derives data from ~1000 points in parallel, while the CMM obtains data serially, limiting the practical number of measured points. Photogrammetry is also attractive because, unlike CMM, it is a non-contact method of measurement. This will be even more important when examining a future window design that is planned to be even thinner.

## 6.6 Initial test setup (Window 4)

The shape of the Window 4 was measured via photogrammetry by imaging the window from both the concave and convex sides, tying both sets of measurements together in one coordinate system via common targets in the referencing fixture.

## 6.7 Comparison to design

The photogrammetric results for Window 4 indicated that there is a bit of eccentricity in the window's shape. This is evident from two approaches to view the data (Figure 18), whisker plots and plots of the difference between measurement and design (Figure 19).

The data indicated a central thickness of 330µm. This value, thinner than indicated by the CMM, is consistent with a burst pressure lower than that predicted by the FEA, which used the CMM thickness of 348um.

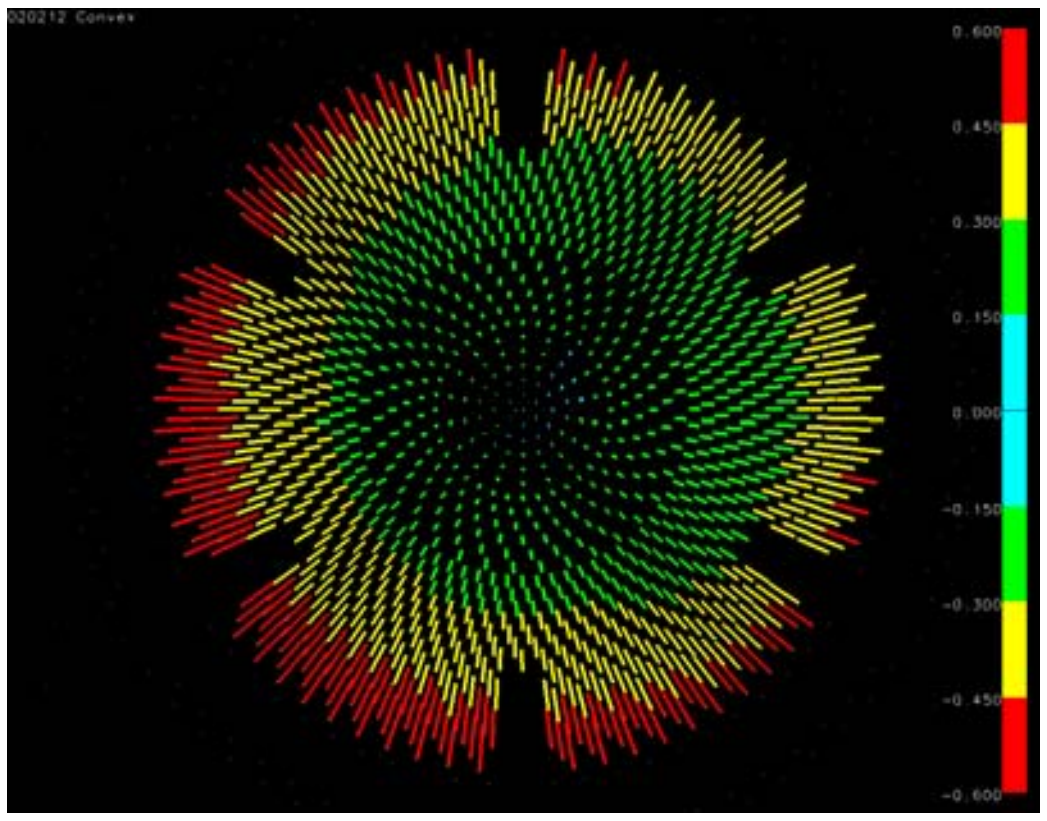

Figure 18

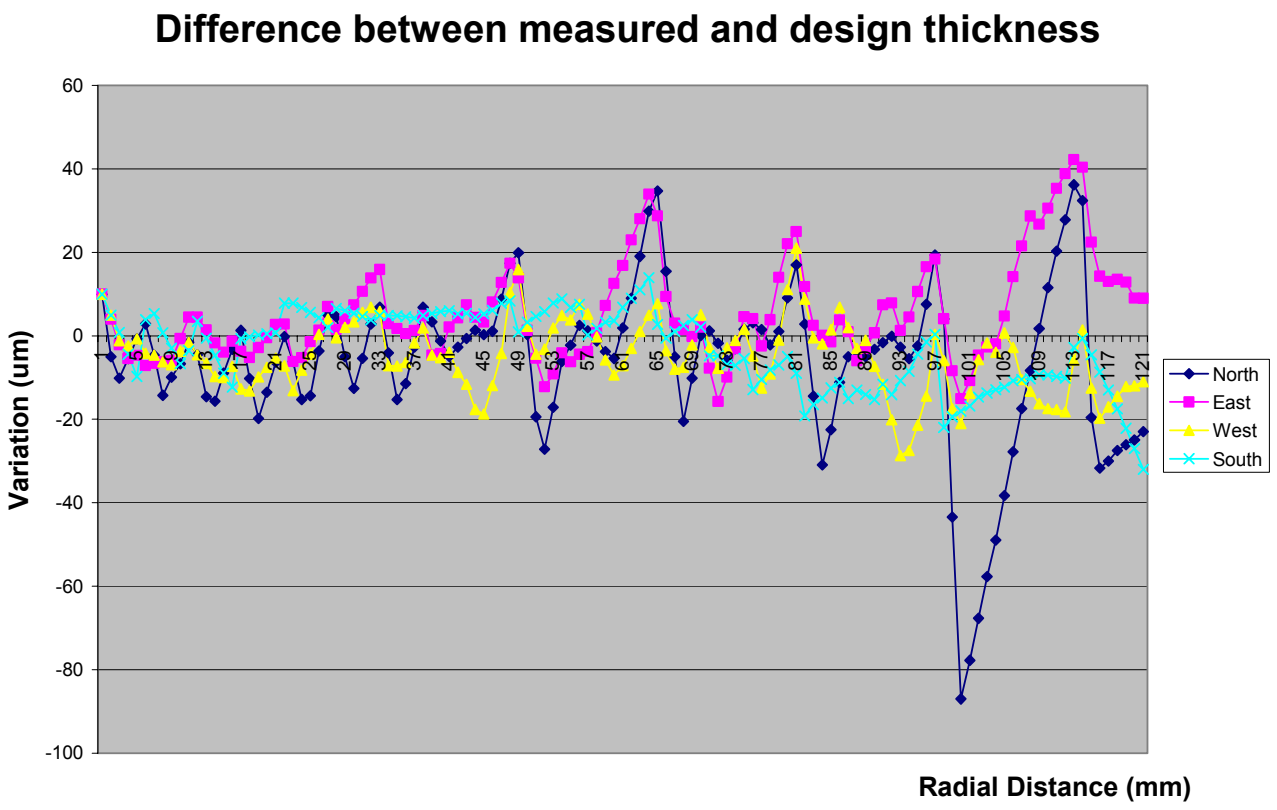

Figure 19

# 7. SURFACE MODELING

Surface modeling techniques are being developed to enhance the application of the photogrammetric technique to material control.

## 7.1 TIN - Triangular Irregular Network

An initial attempt was made to use a TIN to model the surface. A TIN is a series of triangles, similar to that of a geodesic dome, made from the nodal points on a surface (Fig.20). Each triangle, or planar facet, represents a bounded region that approximates the surface. By supplying an X-Y pair within the boundary to the equation of the plane, a value of Z may be determined. Small errors occur when the chosen X-Y pair is away from any of the nodal points. With a triangle measuring approximately 6mm on a side, and a nominal 300mm radius of the spherical surface, the maximum separation between the plane and the sphere is about 6μm, always the same sign. Errors in thickness using TIN modeling for the convex and concave surfaces could be as great as +/- 15μm. Since the target pattern density was at the limit, the use of TIN modeling was abandoned.

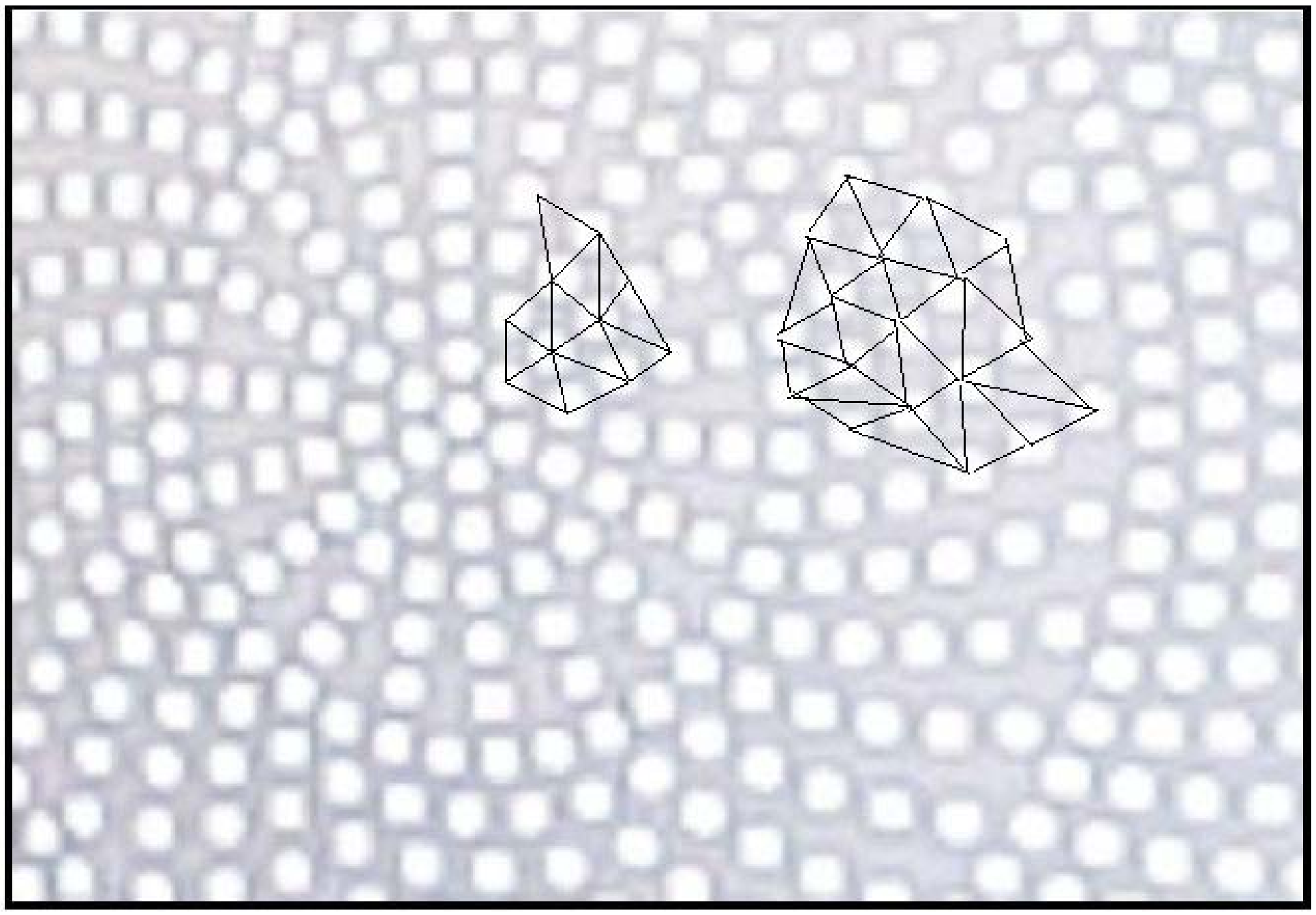

**Figure 20**

### 7.2 Small-Patch Sphere Fit (SPSF)

In an effort to better model the window surface, a technique known as SPSF was developed. Taking advantage of the nearly spherical shape of the window, even at maximum deformation, a solution involving the best-fit sphere determination for a small number of adjacent nodal points,(15 to 25), was suggested. A small number of sample solutions have shown that this is a very promising approach, providing best-fit RMS values in the 2-3μm range, similar to that of the RMS values for the nodal points selected.

## 8. FUTURE IMPROVEMENTS

### 8.1 New window design

A new shape for the absorber window is being developed which seeks an even thinner central region. The surface shape being proposed is called a torisphere, a surface comprised of an outer partial torus, transitioning into a spherical dome. This shape should act much like a bellows, thereby allowing operation at higher pressures for a given thickness (Figure 21).

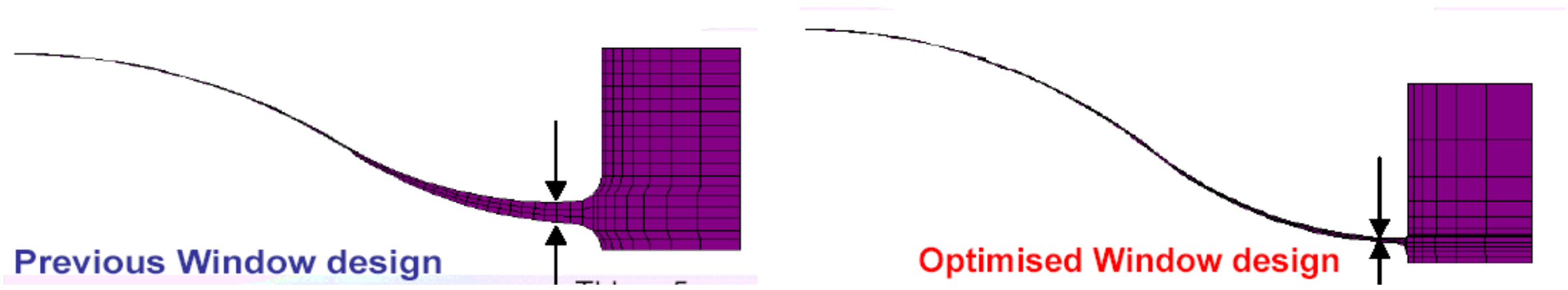

**Figure 21**

### 8.2 Production inspection

Production inspection of the absorber windows will be required. The discussion of issues associated with this activity is continuing. The most likely scenario will call for photogrammetric inspection of each window produced. Whether that will be pressurized dynamic testing, conformance to design shape verification, or some combination of both, is currently being considered.

## 9. CONCLUSIONS

This effort has shown:

- Non-contact projection targeting can be successfully integrated into the digital photogrammetry measurements process for objects of modest (< 0.5m) size, with a precision of 3-5µm
- Repetitive displacement measurements are a suitable substitute for traditional strain gages in determining stress and represents a great advantage where non-contact is crucial
- The process is suitable for both static measurement of shape for conformance to design, as well as dynamic deformation measurements for functional performance of the absorber windows
- Because of coverage, speed, and portability, photogrammetry is particularly well suited for production inspection

## 10. ACKNOWLEDGEMENTS

This work would not have been possible without the guidance of Mary Anne Cummings, the expert engineering and FEA of the window design by Ed Black and Wing Lau, and the careful design and construction of the room temperature and cryogenic pressurization systems by Sasha Dychkant.

### Appendix A - 'Beamstrahlung'

'Beamstrahlung' is synchrotron radiation emitted by a particle as it traverses a bunch of particles traveling in the opposite direction due to the electromagnetic field of the bunch.

### Appendix B - Determination of the length of the astronomical unit

A radar pulse is sent in the direction of Venus, and the time between its transmission and reception is measured. Since time can be measured with great accuracy, the distance to Venus and the dimensions of its orbit can be established within a kilometer. Once repeated measurements of the distance to Venus at closest approach and at most distant separation are acquired, the diameter and eccentricity of both the Earth's and Venus' orbit can be computed. The mean distance from the Earth to the Sun can then be calculated as the mean of these two distances. A check on the Earth-Venus distance has been obtained from trajectories of space vehicles sent to Venus [6].